\documentclass[a4paper,12pt]{article}
\pdfoutput=1 

\usepackage{jheppub} 
\usepackage[export]{adjustbox}
\usepackage[utf8]{inputenc}
\usepackage[compat=1.0.0]{tikz-feynman}

\usepackage{multirow}
\usepackage{soul}

\usepackage{slashed}
\usepackage{bm}

\usepackage{fancyvrb}
\usepackage{fvextra}

\usepackage{comment}

\usepackage{tikz}
\usepackage{tkz-euclide}

\usepackage{booktabs}
\usepackage{epsfig,amsfonts,mathrsfs,amsmath,amssymb,graphicx,color,slashed,multirow}
\usepackage{amsmath,latexsym,amssymb,graphicx,slashed,hyperref,color,enumerate,url,cancel,gensymb}
\usepackage{textcomp}
\usepackage{physics}
\usepackage{textcomp}

\newcommand{\be}{\begin{equation}}
	\newcommand{\ee}{\end{equation}}
\newcommand{\bea}{\begin{equation} \begin{aligned}}
	\newcommand{\eea}{\end{aligned} \end{equation}}

\newcommand{\ba}{\begin{array}}
\newcommand{\ea}{\end{array}}
\def\bV{\begin{Verbatim}[breaklines=true, breakanywhere=true]}
\def\eV{\end{Verbatim}}

\newcommand {\red} 

\def \ie{{\it i.e. }}

\def \GeV{{\, \mathrm{GeV}}}

\newcommand{\al}[1]{\begin{align}\begin{aligned} #1 \end{aligned}\end{align}}

\title{\boldmath 
(In)Visible signatures of the minimal dark abelian gauge sector
}

\author[a]{Ana Luisa Foguel,}
\author[a]{Gabriel M. Salla}
\author[a]{and Renata Zukanovich Funchal}

\affiliation[a]{Departamento de F\'{\i}sica Matem\'atica, Instituto de F\'{\i}sica\\
Universidade de S\~ao Paulo, C. P. 66.318, 05315-970 S\~ao Paulo, Brazil}

\emailAdd{afoguel@usp.br}
\emailAdd{gabriel.massoni.salla@usp.br}
\emailAdd{zukanov@if.usp.br}

\abstract{In this paper we study the present and future sensitivities of the rare meson decay facilities KOTO, LHCb and Belle II to a light dark sector of the minimal dark abelian gauge symmetry where a dark Higgs $S$ and a dark photon  $Z_D$ have masses $\lesssim 10$ GeV. We have explored the interesting scenario where $S$ can only decay to a pair of $Z_D$'s and so contribute to visible or invisible signatures, depending on the life-time of the latter. Our computations show that these accelerator experiments can access the dark Higgs (mass and scalar mixing) and the dark photon (mass and kinetic mixing) parameters in a complementary way. We have also discussed how the CMS measurement of the SM Higgs total decay width and their limit on the Higgs invisible branching ratio can be used to extend the experimental reach to dark photon masses up to $\sim 10$ GeV, providing at the same time sensitivity to the gauge coupling associated with the broken dark abelian symmetry. 
}

\begin{document}

\maketitle

\section{Introduction}
\label{sec:intro}

Despite the enormous success of the Standard Model (SM) in describing the interactions of elementary particles, it still fails to give an explanation to neutrino masses, dark matter and dark energy. In light of all null results from the LHC in the search of Beyond the Standard Model (BSM) physics, feebly interacting dark sectors became one of the most well motivated extensions of the SM. Out of all such extensions, the addition of a new massive vector, named here dark photon, associated to a new $U(1)_D$ gauge symmetry, has strong theoretical motivations and also offers a rich phenomenology. If we further assume that no state is charged under this new $U(1)_D$, this model is controlled by only two parameters: the mass of the dark photon and its kinetic mixing $\epsilon$ to the hypercharge gauge field. For recent reviews on the subject and a summary of the present experimental bounds, we refer the reader to \cite{Fabbrichesi:2020wbt,Graham:2021ggy}.

In spite of its compelling simplicity, the model of a massive dark photon is nothing but an effective theory due to the absence of a mass generation mechanism and bad ultraviolet (UV) behavior of the longitudinal modes \footnote{Invoking a Stückelberg mass might avoid problems in this regard, however, as indicated in \cite{Kribs:2022gri}, there are other allowed operators that could spoil the model in the UV.}. To take this into account, we must modify the dark photon model by either embedding it into an Effective Field Theory or by directly UV completing it. The first scenario assumes that all other BSM states that might be charged under $U(1)_D$ are very heavy and one can thus include their effects in the low energy theory by adding higher dimensional operators. The impacts of the dimension 6 operators to the dark photon phenomenology was first investigated in \cite{Rizzo:2021lob, Barducci:2021egn}, and it was found  that under certain circumstances the effective operators can dramatically change the present bounds on the dark photon parameter space. In the second scenario the theory is complemented by a scalar sector that spontaneously breaks the dark $U(1)_D$ and therefore gives  mass to the dark photon. The simplest realization of this mechanism, in which the new scalar $S$ is a SM singlet, is known in the literature as the Hidden Abelian Higgs Model (HAHM) \cite{Curtin:2014cca}. Interestingly, the HAHM  is a particular combination of the scalar and vector renormalizable portals with the addition of an interaction between the dark Higgs $S$ and the dark photon due to  gauge interactions. Previous studies show that the standard experimental constraints on the dark Higgs and dark photon parameter spaces are also drastically modified in this case \cite{Curtin:2014cca, CMS:2022yoy,ATLAS:2022bll,CMS:2021sch,Elkafrawy:2021mrm,CMS:2021pcy,ATLAS:2021ldb,Ferber:2022ewf, Araki:2020wkq,2012.02538}.

In this paper we further explore how the phenomenology of the dark Higgs and of the dark photon in the context of the HAHM affect 
experimental observables, focusing on the situation where \textit{both} particles are light ($\lesssim 10$ GeV). More precisely, we are interested in novel meson decay signatures involving 4 charged leptons in the final state, which in this model can take place through the gauge connection of scalar and vector portals. In the search for these signatures, one can benefit  
from the future prospects of experiments at the \textit{intensity} and \textit{high precision} frontiers. In particular, the KOTO \cite{Yamanaka:2012yma}, LHCb \cite{LHCb:2008vvz} and Belle II \cite{Belle-II:2018jsg,Belle-II:2010dht} experiments aim to probe, respectively, extremely rare kaons, $B$-mesons and $\Upsilon$'s
decays with increasing luminosity in the years to come. 
Their data, as we show here, can be used to search for the HAHM 
signatures in meson decays leading to potential discovery 
or to  stringent constraints on the HAHM parameter space.
Apart from rare meson decays, the scalar-vector connection can deeply affect the SM-like Higgs boson phenomenology due to the mixing with the dark scalar. Since the new dark sector will be taken to be much lighter than the SM-like Higgs, the latter can decay into dark particles and thus contribute to its invisible width. As we are going to see, we can obtain a set of conditional constraints from requiring the invisible branching ratio to be consistent with 
experimental bounds.

The paper is organized as follows. In section~\ref{sec:theo} we review the most important theoretical aspects of the model, giving particular emphasis to the connection between the scalar and vector portals. Section~\ref{sec:limHad} is dedicated to the analysis of KOTO, LHCb and Belle II experiments using the relevant meson decays. Then, in section~\ref{sec:HiggsInv}, we show how the bounds from these experiments can be complemented by constraints coming from measurements of the Higgs invisible branching ratio in CMS. We conclude in section~\ref{sec:conc}. We have two appendices, in appendix \ref{app:HAHM} we give a brief summary of the HAHM and show explicitly the expressions for the relevant decay widths, and in appendix \ref{app:BtoK} we collect the expressions for the matrix elements we used in sections \ref{sec:lhcb}.

\section{The HAHM model}
\label{sec:theo}

\subsection{Theoretical framework}

The model known as HAHM~\cite{Curtin:2014cca} consists of extending the SM gauge content by an extra $U(1)_{D}$ in the simplest UV complete way. This additional abelian gauge group is associated with a new neutral vector boson $Z_D$, the dark photon, which acquires mass due to the vacuum expectation value (vev) $v_S$ of a new SM scalar singlet $S$. Due to this, the physical particle associated with this field will be referred to as the dark Higgs. While the SM particles are uncharged under the new symmetry, $S$ can take an arbitrary charge $q_S$, which from now on we fix to be 1. In this setup, the SM Lagrangian is modified in both the scalar and neutral gauge sectors and, as a consequence, we end up having both portals to the hidden sector.

At very high energies, before any spontaneous symmetry break takes place, the dark abelian gauge boson $\hat Z_{D\mu}$ mixes kinematically with the hypercharge gauge boson $\hat B_\mu$ through the following Lagrangian
\begin{equation} \label{eq:KMlag}
    \mathcal{L}_{ZB} = - \frac{1}{4} \hat B_{\mu \nu} \hat B^{\mu \nu}  - \frac{1}{4} \hat Z_{D \mu \nu} \hat Z_D^{\mu \nu} + \frac{\epsilon}{2 c_W} \hat Z_{D \mu \nu} \hat B^{\mu \nu},
\end{equation}
where $\hat B_{\mu \nu}$ and $\hat Z_{D\mu \nu}$ are, respectively, the $U(1)_Y$  and $U(1)_D$ field strength tensors, $\epsilon$ parameterizes their kinetic mixing and $c_W \equiv \cos{\theta_W}$ is the cosine of the weak mixing angle. The hatted fields indicate states with non-canonical kinetic terms. At the same time, in the scalar sector, the dark scalar $S$ interacts with the SM-like doublet $H$ via a quartic term in the potential. The Lagrangian that contains this latter interaction reads
\begin{equation}\label{eq:L_scalar}
    \mathcal{L}_{\mathrm{scalar}} = |D_\mu H|^2  + |D_\mu S|^2 - V(H,S) \,,
\end{equation}
where the scalar potential $V(H,S)$ is given by
\begin{equation}\label{eq:scalar_pot}
     V(H, S) = -\mu^2 |H|^2 + \lambda |H|^4 - \mu_S^2 |S|^2 + \lambda_S |S|^4 + \kappa |H|^2 |S|^2,
\end{equation}
with $\kappa$ parametrizing the scalar mixing and we require that $\mu^2,\mu_S^2,\lambda,\lambda_S>0$ in order to spontaneously break both $SU(2)_L\times U(1)_Y$ and $U(1)_{D}$. The SM-like Higgs and the dark Higgs covariant derivatives are
\begin{equation}\label{eq:HcovD}
    D_\mu H = \partial_\mu H - i g W_{a \mu} \tau^a H - \frac{1}{2} i g' \hat B_\mu H \,,
\end{equation}
\begin{equation}\label{eq:ScovD}
    D_\mu S  = \partial_\mu S  - i  g_D  \hat Z_{D \mu} S \,,
\end{equation}
where $g$, $g'$ and $g_D$ are the $SU(2)_L$, $U(1)_Y$ and $U(1)_D$ gauge couplings, respectively.

After spontaneous symmetry breaking, kinetic and mass diagonalizations, we recover the physical states: $Z$, $Z_D$, $h$ and $S$ \footnote{We make a slight abuse of notation and use $S$ to denote both the complex scalar before spontaneous symmetry breaking as well as the physical dark Higgs. The discrimination between the two should be clear from the context.}. In the gauge sector, for $\epsilon\ll 1$ and $m_{Z_D}\ll m_Z$, the $Z$-boson state is  almost entirely given by the un-diagonalized SM $Z^0$-boson with only a small mixing with the dark vector. Furthermore, the mass of the physical dark photon will be approximately expressed as $m_{Z_D}=g_D v_S$ while its couplings to the SM fermions will take place via the kinetic mixing, \textit{i.e.} in a photon-like manner with the fermion charge substitution $q^f_{\rm em}e \to \epsilon q^f_{\rm em}e$. Similarly in the scalar sector, for $\kappa\ll 1$ and $m_S \ll m_h$, the masses of the physical states are written as
\bea\label{eq:scalar_masses}
m_h^2 & \simeq 2\lambda v^2 (1+s_h^2) - s_h^2 2\lambda_S v_S^2,\\
m_S^2 & \simeq 2\lambda_S v^2_S (1+s_h^2) - s_h^2 2\lambda v^2,
\eea
where we have defined the scalar mixing angle $\sin \theta_h \equiv s_h$ as
\be
s_h \simeq \frac{\kappa v v_S}{m_S^2-m_h^2}+\order{\kappa^2}.
\ee
Given that in eq.~\eqref{eq:scalar_masses} only $s_h$ appear instead of $\kappa$, we will focus on the former and express all physical observables in terms of it. In particular, we will assume that $s_h\ll 1$. Thus, for a small mixing, we conclude that the SM Higgs $h$ is mostly constituted by the un-mixed Higgs $h_0$ with a small contamination of the un-mixed dark scalar $S_0$. As a direct consequence of this mixing, the Higgs (dark Higgs) inherits all interactions involving $S_0$ ($h_0$)  with an extra $s_h$ suppression. Similarly, all SM-like Higgs physical couplings get suppressed by $c_h\equiv\cos \theta_h$ \footnote{Measurements of the Higgs coupling strength imposes that $c_h\gtrsim 0.9$, which in turn implies that $s_h\lesssim 0.1$ \cite{Workman:2022ynf}.}.

It is worth remarking that the Lagrangians in eqs.~\eqref{eq:KMlag} and \eqref{eq:L_scalar} represent, respectively, a particular combination of the standard vector and scalar renormalizable extensions of the SM. What makes this model special is the fact that the dark scalar interacts with the dark vector through the covariant derivative \eqref{eq:ScovD}. Consequently, the modifications from the usual phenomenology of the vector and scalar portals will strongly depend on this dark gauge coupling. In order to quantify such changes, we will explore in this paper the regime in which the dark gauge interaction dominates the dynamics of the dark Higgs.

One last remark about the HAHM is in order. As it will be shown in sections \ref{sec:limHad} and \ref{sec:HiggsInv}, the sensitivity of the experiments considered here can reach regions in parameter space where $\epsilon\ll s_h$. Such a large parametric separation between the two couplings is not very natural, as the kinetic mixing parameter can receive contributions from two-loop diagrams that are proportional to $s_h$. In order to achieve $\epsilon\ll s_h$ we assume some degree of fine-tuning of the parameters.

\subsection{Decay Widths of the Dark Sector}

%
\begin{figure}[t]
\begin{center}
\includegraphics[width=0.8\textwidth]{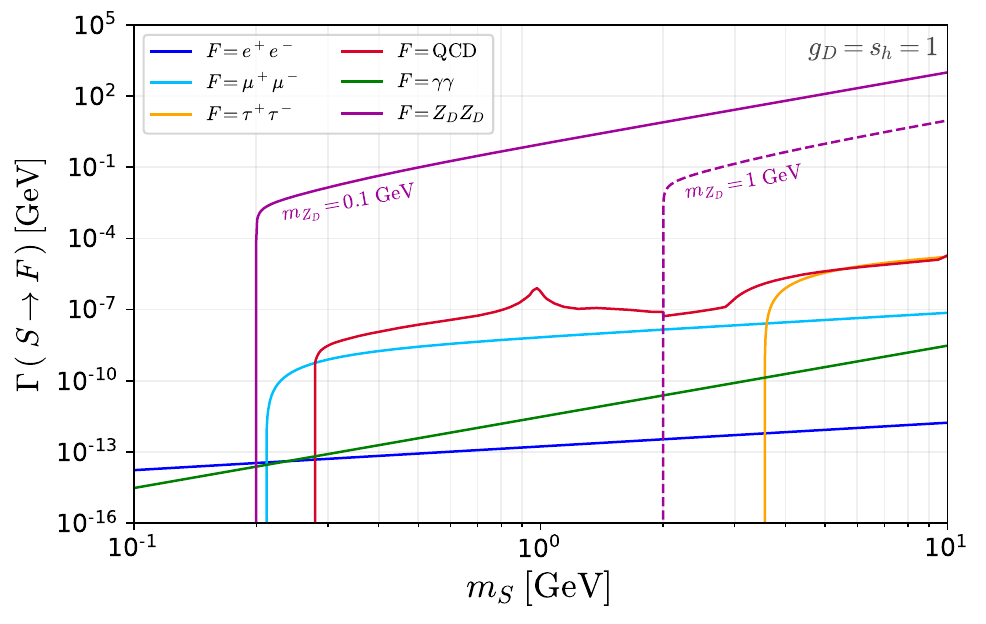}
\end{center}
\vglue -0.8cm
\caption{\label{fig:DarkHiggsW} Dark Higgs partial decay widths into electrons (dark blue), muons (light blue), taus (orange), light hadrons, quarks and gluons (dark red), photons (green) and dark photons (purple) with $m_{Z_D}= 0.1 \GeV$ (solid) and $m_{Z_D}= 1 \GeV$ (dashed). We fixed the couplings to $g_D = 1$ and $s_h=1$.}
\end{figure}
%

%
\begin{figure}[t]
\begin{center}
\includegraphics[width=0.45\textwidth]{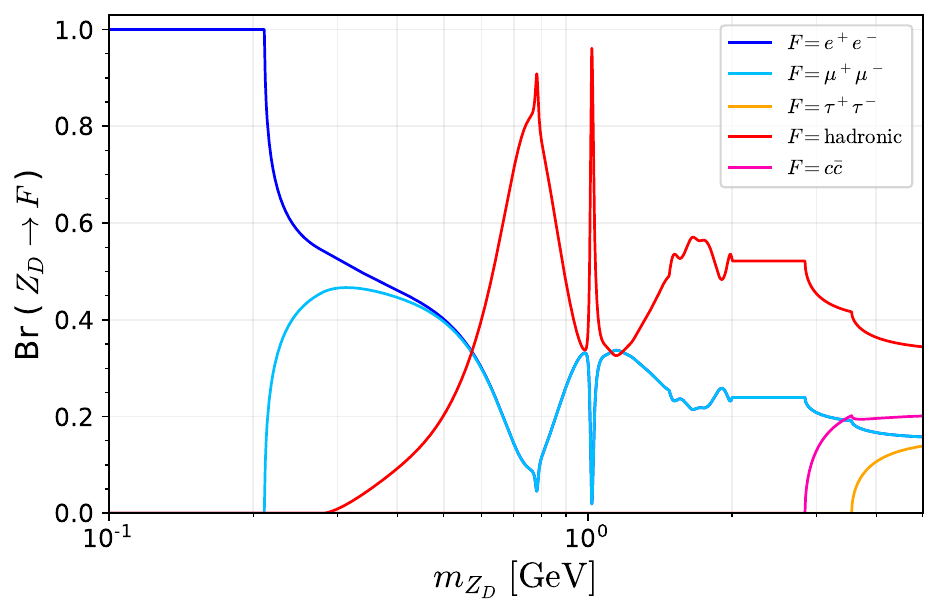}
\includegraphics[width=0.45\textwidth]{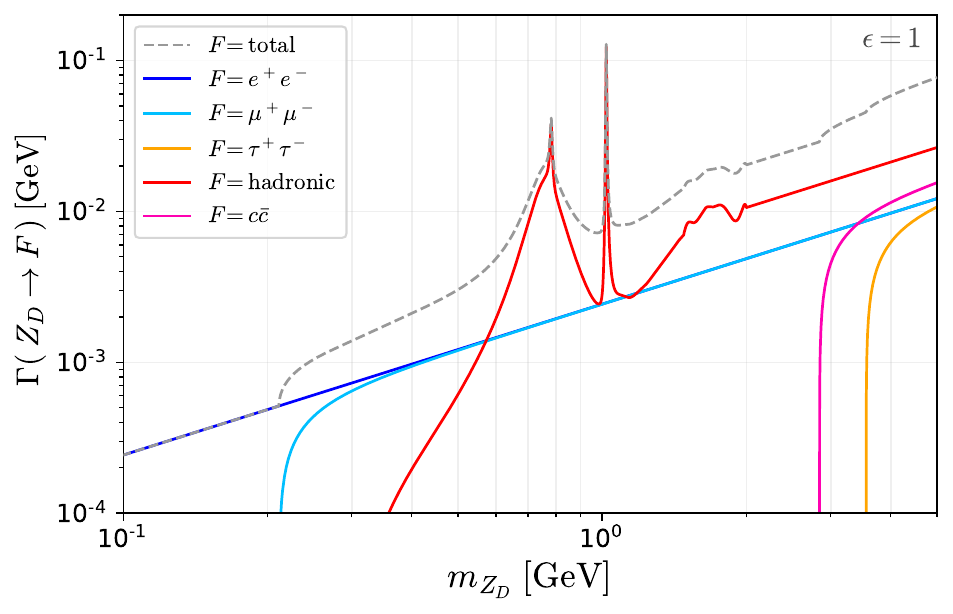}
\end{center}
\vglue -0.8cm
\caption{\label{fig:DarkWidths} The left panel shows the branching ratio of the dark photon into electrons (dark blue), muons (light blue), taus (orange), hadrons and light quarks (red) and $c \bar c$ (pink). The right panel shows the dark photon total (dashed) and partial (solid) decay widths for a fixed kinetic mixing parameter $\epsilon=1$.}
\end{figure}

After spontaneous symmetry breaking, the dark Higgs covariant derivative in 
eq.~\eqref{eq:L_scalar} contains the following term
\be\label{eq:dark_gauge_connection}
\mathcal{L}_{\mathrm{scalar}}\supset g_D m_{Z_D} S Z_D^\mu Z_{D\mu},
\ee
where we took $c_h\simeq 1$. If $m_S \geq  2\, m_{Z_D}$, the corresponding decay width reads
\be\label{eq:sZDZD}
\Gamma(S \to Z_D Z_D) =  \left(\frac{g_D}{m_{Z_D}}\right)^2 \frac{1}{32 \pi \, m_S} \left(m_S^4 - 4\, m_S^2 m_{Z_D}^2+ 12 \, m^4_{Z_D}\right)\sqrt{1-   \frac{4m_{Z_D}^2}{m_S^2}}\, .
\ee
All other possible decay channels of $S$ into SM states must proceed via the mixing with the Higgs and are therefore suppressed by $s_h^2$. We give the formulas for other decay widths in appendix~\ref{app:HAHM}. Since the width to dark photons is independent of $s_h^2$, there is a particular regime where $\Gamma(S \to Z_D Z_D)\ll \Gamma_S(\text{total})$, thus reproducing the usual dark Higgs phenomenology, and the opposite regime $\Gamma(S \to Z_D Z_D)\simeq \Gamma_S(\text{total})$ in which the gauge sector dominates. For the latter scenario to hold, the condition
\be\label{eq:condition_sZDZD}
g_D\gg 7\cdot 10^{-3} s_h\, ,
\ee
must be satisfied, where we have used that $m_S \geq  2\, m_{Z_D}$. In figure~\ref{fig:DarkHiggsW} we show the partial decay widths of the dark Higgs as a function of its mass, fixing $s_h=g_D=1$. In particular, we highlight the Higgs-inherited contributions from charged leptons, photons and from QCD\footnote{Here QCD denotes hadrons when $m_S\lesssim 2$ GeV and quarks and gluons for higher masses. The low-energy width to hadrons was taken from \cite{Winkler:2018qyg}.} that scale with $s_h^2$, and also show the width \eqref{eq:sZDZD} for two dark photon masses. In the figure, we can clearly see the dominant behavior of $\Gamma(S\to Z_DZ_D)$.

Fixing the mass hierarchy $m_S \geq  2\, m_{Z_D}$, the partial widths of the dark photon will not be affected by the scalar sector and will be thus described entirely by the kinetic mixing. The decay rate to fermions is then
\begin{equation}\label{eq:ZDff}
    \Gamma(Z_D \to \bar f f)=N_c\frac{(q^f_{\rm em} e \epsilon)^2}{12 \pi} \, m_{Z_D} \left( 1+  \frac{2m^2_f}{m_{Z_D}^2}\right)\sqrt{1-  \frac{4m_f^2}{m_{Z_D}^2}}\, ,
\end{equation}
where the number of colors $N_c=3\, (1)$ for quarks (leptons), $q^f_{\rm em}=-1$ for charged leptons and $q^f_{\rm em}=2/3 \, (-1/3)$ for $u$-type ($d$-type) quarks. In the range $0.2 \lesssim m_{Z_D}/{\rm GeV} \lesssim 2$, the decays of $Z_D$ to hadrons are extremely relevant\footnote{For a recent evaluation of these hadronic decays, see for instance ref.~\cite{Foguel:2022ppx}.}. Above $m_{Z_D} \sim 2$ GeV, the perturbative QCD contribution, at first order given by eq.~(\ref{eq:ZDff}), dominates.

On the left panel of figure~\ref{fig:DarkWidths} we show the branching ratio of $Z_D$ to charged fermions, hadrons 
and $c\bar c$ as a function of $m_{Z_D}$. For $m_{Z_D} \lesssim 0.5$ GeV, $Z_{D}$ will mainly decay into a pair of charged leptons (electrons or muons), above this mass hadronic decays start to dominate but, except for the masses that correspond to hadronic resonances, leptonic decays can still make up for 10 to 30\% of the total $Z_D$ decays. On the right panel of figure~\ref{fig:DarkWidths} we show the partial and total 
decay width of the dark photon as a function of $m_{Z_D}$
for $\epsilon=1$.

\section{Limits from Hadronic Decays}
\label{sec:limHad}

We will be discussing here exotic meson decays involving dark particles in three experiments: the rare  kaon decay experiment KOTO, LHCb and the $B$-factory Belle~II. More precisely, we will be interested in observing how different is the HAHM phenomenology when compared to the one of the usual scalar and vector portals.

\begin{figure}[t]
\centering
\begin{tikzpicture}
\begin{feynman}
\vertex (M1) {$M$};
\vertex[right=1.5cm of M1] (M2);
\vertex[right=1.5cm of M2] (Ma);
\vertex[above=0.8cm of Ma] (Ml) {$M'$};
\vertex[below=0.5cm of Ma] (s1);
\vertex[right=1.2cm of s1] (sa);
\vertex[above=1.0cm of sa] (zp1);
\vertex[below=1.0cm of sa] (zp2);
\vertex[right=1.0cm of zp1] (zp1a);
\vertex[above=0.3cm of zp1a] (l11) {$\ell^-$};
\vertex[below=0.3cm of zp1a] (l12) {$\ell^+$};
\vertex[right=1.0cm of zp2] (zp2a);
\vertex[above=0.3cm of zp2a] (l21) {$\ell^-$};
\vertex[below=0.3cm of zp2a] (l22) {$\ell^+$};
\diagram*{
(M1) -- (M2) -- (Ml),
(M2) -- [scalar,edge label'=$S$] (s1),
(s1) -- [photon,edge label=${\scriptstyle Z_D}$] (zp1),
(s1) -- [photon,edge label'=${\scriptstyle Z_D}$] (zp2),
(zp1) -- [fermion] (l11),
(zp1) -- [anti fermion] (l12),
(zp2) -- [fermion] (l21),
(zp2) -- [anti fermion] (l22),
};
\end{feynman}
\end{tikzpicture}
\caption{Representative diagram for the process we consider in our analysis of a meson $M$ decaying to a SM state $M'$ and a dark Higgs $S$, that subsequently decays to a pair of dark photons. See text for more details.}
\label{fig:diagram}
\end{figure}
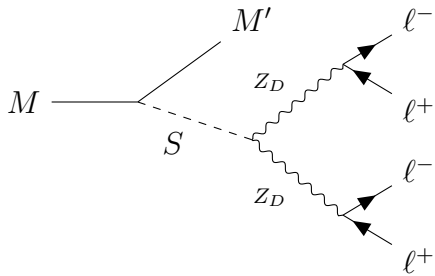

In section~\ref{sec:theo} we saw that the HAHM presents the feature of connecting both renormalizable portals via the dark gauge interaction given in eq.~\eqref{eq:dark_gauge_connection}. This interaction can contribute to the dark Higgs decay width, as evidenced by eq.~\eqref{eq:sZDZD}, and can even completely dominate the total width if the condition stated in eq.~\eqref{eq:condition_sZDZD} is met. Since many searches for light scalars rely on Higgs-like decays that are suppressed by $s_h$ \cite{Winkler:2018qyg,LHCb:2015nkv,LHCb:2016awg,Blumlein:1991xh,Belle:2021rcl}, having $\text{BR}\left(S\to Z_DZ_D\right)\simeq 1$ will deeply impact such searches. For the following analysis of KOTO, LHCb and Belle II, we will assume the condition given in eq.~\eqref{eq:condition_sZDZD} to hold and thus that the dark Higgs decays $100\%$ to a pair of dark photons. In this regime, each dark photon will afterwards decay to SM particles, which for our visible analysis we take to be a pair of charged leptons. Therefore, in a similar spirit to \cite{0911.4938,Hostert:2020xku}, we consider novel signatures in meson decays with four leptons in the final state (see figure~\ref{fig:diagram} for a representative diagram).

If the experimental signal is visible, \textit{i.e.} the four leptons are being measured, the number of events at each experiment is given by
\be\label{eq:Nevts}
N_\text{evt} = N_M ~{\rm BR}\left(M\to S+ M'\right) ~P_\text{dec}\left(l_\text{in}\to l_\text{out}\right)~ f_\text{geom}~ {\rm BR}\left(Z_D\to \ell^-\ell^+\right)^2 ~\varepsilon ,
\ee
where $N_M$ is the initial number of mesons produced, ${\rm BR}\left(M\to S+ M'\right)$ is the branching ratio of  the meson $M$ to decay into $S$ plus another SM state $M'$ and $P_\text{dec}\left(l_\text{in}\to l_\text{out}\right)$ is the total decay probability with $l_{\rm in, \rm out}$ the distances at which the particles enter and exit the detector, respectively. In eq.~\eqref{eq:Nevts} we also consider the geometrical acceptance $f_{\rm geom}$ of the detector and the efficiency $\varepsilon$ of detection. The decay probability above takes into account the fact that the dark particles can be long-lived and travel macroscopic distances. For the case of the dark photon the decay length is controlled by its mass and the kinetic-mixing parameter $\epsilon$, and can typically be long-lived for $\epsilon \lesssim 10^{-4}$. For the dark Higgs instead, under the assumption that $\Gamma_S({\rm total}) \simeq \Gamma(S\to Z_DZ_D)$, the decay length is dictated by the dark gauge coupling $g_D$. Considering not too small values of $g_D, \textit{i.e.}$ $g_D\gtrsim 10^{-3}$, and $m_S\gtrsim \order{0.1~{\rm GeV}}$, we can guarantee that the dark Higgs decays promptly and the number of events will thus not depend on its decay length. In this situation we can rewrite the decay probability as
\be\label{eq:ProbDec}
P_\text{dec}\left(l_\text{in}\to l_\text{out}\right) = P_\text{dec}^{Z_D,1}P_\text{dec}^{Z_D,2},
\ee
where each $P_\text{dec}^{Z_D,i}$ is the probability for the $i$th dark photon to decay inside the detector and has the following expression
\be\label{eq:ProbDec_DP}
P_\text{dec}^{Z_D,i} = e^{-l_\text{in}^i/\lambda^i}-e^{-l_\text{out}^i/\lambda^i},
\ee
with $\lambda^i$ the corresponding decay length. If on the other hand the experiment relies on invisible signatures to constraint New Physics, we must guarantee that the dark particles escape the detector rather than falling inside of it. Still considering that the dark Higgs decays promptly, this condition is translated to having both dark photons escaping the detector, meaning that the decay probability of eq.~\eqref{eq:ProbDec_DP} is modified to
\be
\label{eq:ProbDec_DP_inv}
P_\text{dec}^{Z_D,i} 
\to e^{-l_\text{out}^i/\lambda^i}.
\ee
In order to consider an invisible signal we must further modify eq.~\eqref{eq:Nevts} by taking ${\rm BR}\left(Z_D\to \ell^-\ell^+\right)\to 1$.

One very important property of the number of events is that, when considering that $S$ decays solely to dark photons and promptly, we achieve a de-correlation between production and the decay probability. As we will see more explicitly, the production branching ratio of $S$ will depend on the mixing angle $s_h^2$, while the decay probability will depend on $\epsilon$ because only the dark photons might be long-lived. Hence, production and detection depend on different sets of parameters and so on different couplings. This will allow us to observe drastic changes on the present and future experimental sensitivity to the model. We again emphasize that this is only possible due to the dark gauge interaction shown in eq.~\eqref{eq:dark_gauge_connection}.

\subsection{KOTO}

KOTO is an experiment at the Japan Proton Accelerator Research Complex (J-PARC)~\cite{Yamanaka:2012yma}  dedicated to studying the CP-violating  rare decay $K_L \to \pi^0 \nu \bar\nu$ aiming to measure the SM predicted branching fraction of $(3.00 \pm 0.30) \times 10^{-11}$~\cite{Buras:2015qea}. 

The $K_L$ beam is produced by colliding 30 GeV protons from J-PARC Main Ring accelerator with a gold production target. The measured flux  at the exit of the beam line is $2.1 \times 10^{-7}$ $K_L$'s per  protons on target (POT)~\cite{KOTO:2020prk}, with the peak $K_L$ momentum being 1.4 GeV, while a total of $3.05 \times 10^{19}$ POT was collected from 2016 to 2018. The main background events to the $K_L \to \pi^0 \nu \bar\nu$ signal were estimated to be $K^{\pm} \to \pi^0 e^\pm \nu$, $K_L \to 3 \pi^0$ and beam-halo $K_L \to 2 \gamma$, contributing to a total of $1.22\pm 0.26$ expected background events for the total data set~\cite{KOTO:2020prk}. The collaboration observed 3 events in the signal region, which is consistent with the expected background allowing them to place the bound ${\rm BR}(K_L \to \pi^0 \nu \bar\nu)< 4.9 \times 10^{-9}$ at 90\% CL~\cite{KOTO:2020prk}. Acccording to ref.~\cite{Liu:2020qgx}, the same number of $K_L$ decays considered in this analysis as well as the same estimated background applies for $K_L \to \pi^0 X$, with $X$ a neutral stable state, however, the acceptance is not the same. Taking into account the different acceptance they obtained  the bound ${\rm BR}(K_L \to \pi^0 X)< 3.7 \times 10^{-9}$ at 90\% CL for a massless $X$.

\begin{figure}[t]
\begin{center}
\includegraphics[width=0.8\textwidth]{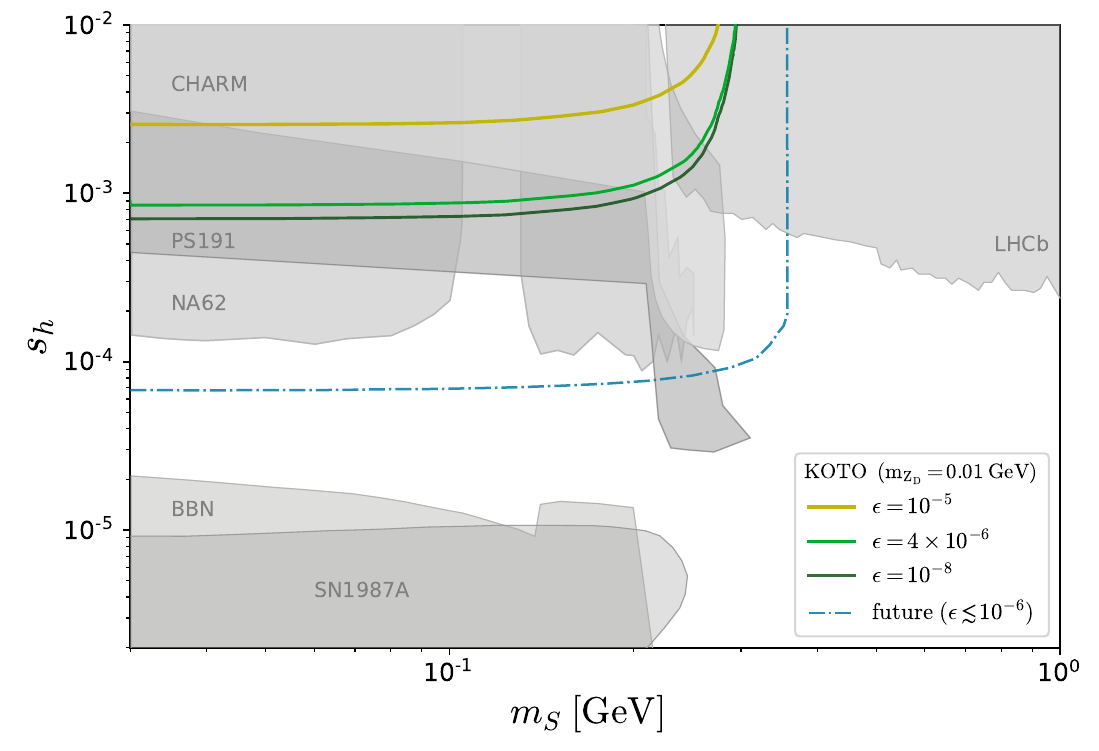}
\end{center}
\vglue -0.8cm
\caption{\label{fig:KOTOlim} Current bound on the HAHM dark Higgs parameter space considering KOTO upper limit ${\rm BR}(K_L \to \pi^0 X)$~\cite{Liu:2020qgx} and fixing $m_{Z_D}=0.01$ GeV. The different solid curves consider $\epsilon=10^{-5}$ (yellow), $4\times  10^{-6}$ (green) and $10^{-8}$ (dark green). We show in blue dashed-dotted the maximum future sensitivity of the KOTO experiment assuming the SM prediction for ${\rm BR}(K_L \to \pi^0 \nu \bar\nu)$ can be attained. The gray regions represent the limits from the LHCb collider experiment ($B \to K S$ search)~\cite{LHCb:2015nkv,LHCb:2016awg}, from proton beam dump experiments CHARM~\cite{Winkler:2018qyg} and PS191~\cite{Gorbunov:2021ccu}, from the fixed-target experiment NA62~\cite{NA62:2021zjw}, as well as  the SN1987A~\cite{Winkler:2018qyg} and BBN~\cite{Fradette:2017sdd} constraints.}
\end{figure}

We can translate this bound on a limit on the mixing $s_h$ as a function of $m_S$, using the branching ratio of $K_L$ decaying to a dark Higgs
\be{\label{eq:KLpis1}} {\rm BR}(K_L \to \pi^0 S)=\frac{s_h^2}{\Gamma_{K_L}} \frac{2\, p_S^{K\pi}}{m_{K}}\frac{ \vert {\cal M}_K\vert^2}{16 \pi\,  m_{K}}\, ,
\ee
where $p_S^{K\pi}=\lambda^{1/2}(m_K^2,m_\pi^2,m_S^2)/(2 \, m_K)$ with $\lambda(a,b,c) = a^2+b^2 +c^2 - 2 (ac+ab+bc)$ and the matrix element ${\cal M}_K$, calculated in ref.~\cite{Leutwyler:1989xj}, is 
\be{\label{eq:MEKLpis}} 
{\cal M}_K = -\frac{m^2_{K}}{v}\left[\frac{7}{18}\gamma_1 \left(1 - \frac{m_S^2-m_\pi^2}{m^2_{K}}\right) -\frac{7}{9} \gamma_2 +\frac{1}{2}\frac{3 \sqrt{2}\, G_F}{16 \pi^2}\sum_{i=u,c,t}V^*_{id}\, m_i^2\, V_{is}\right]\,,
\ee
where $m_K$ and $\Gamma_{K_L}$ are, respectively, the neutral kaon mass  and the $K_L$ total decay width, $m_\pi$ is the neutral pion mass, $G_F$ is the Fermi constant, $V_{id,is}$ and $m_i$ ($i=u,c,t$) are the CKM matrix elements and quark masses. The dominant contribution comes from the quark-$W$ loop diagram  given by last term of the amplitude. According to ref.~\cite{Leutwyler:1989xj}, the parameter $\gamma_1$ can be extracted from data, resulting in $\gamma_1 \simeq 3.1 \times 10^{-7}$, while $\gamma_2 \ll \gamma_1$ and so it will be neglected here. The effect of the $\gamma_1$ term amounts to decrease the main contribution by $\sim 14\%$. Considering all this we may re-write eq.~\eqref{eq:KLpis1} as
\be{\label{eq:KLpis}} {\rm BR}(K_L \to \pi^0 S)\approx 7.5 \times 10^{-3}\, s_h^2 \,  \frac{2\, p_S^{K\pi}}{m_K}\, ,
\ee
where the numerical pre-factor can vary between 7.3 and 7.8 due to the weak dependence on $m_S$.
Since we are in the regime in which the dark Higgs decays promptly to a pair of dark photons, the experimental constraint on ${\rm BR}(K_L\to \pi^0 X)$ does not apply directly to ${\rm BR} (K_L\to \pi^0 S)$, but rather to the following effective branching ratio
\be\label{eq:KOTO_BR_eff}
{\rm BR}_{\rm eff} \equiv {\rm BR} (K_L\to \pi^0 S)\, {\rm BR}(S\to Z_DZ_D)\,  P_{\rm dec}^{Z_D,1}P_{\rm dec}^{Z_D,2},
\ee
where $P_{\rm dec}^{Z_D,i}$ is the probability of the dark photon to decay outside the KOTO detector, given by eq.~\eqref{eq:ProbDec_DP_inv}. The quantity above takes into account the fact that the dark photons must necessarily escape in order to contribute to the signal searched by KOTO. We then see from 
eq.~\eqref{eq:KOTO_BR_eff} that, for given values of $m_{Z_D}$ and $\epsilon$, we can set limits on the $m_S\times s_h$ plane by requiring that ${\rm BR}_{\rm eff} \leq {\rm BR}(K_L\to \pi^0 X)$.

To estimate the effective branching ratio of eq. (\ref{eq:KOTO_BR_eff}) we simulate a $K_L$ flux according to ref. \cite{KOTO:2012zpk} with approximately $6.4\cdot10^{12}$ kaons produced and consider the decay volume used in refs. \cite{Liu:2020qgx,KOTO:2020prk}. The geometry of the KOTO detector is implemented in \verb!MadDump!~\cite{Buonocore:2018xjk}, that, together with the initial $K_L$ flux, allowed us to compute the decay probabilities of the dark photons.

\begin{figure}[t]
\begin{center}
\includegraphics[width=0.7\textwidth]{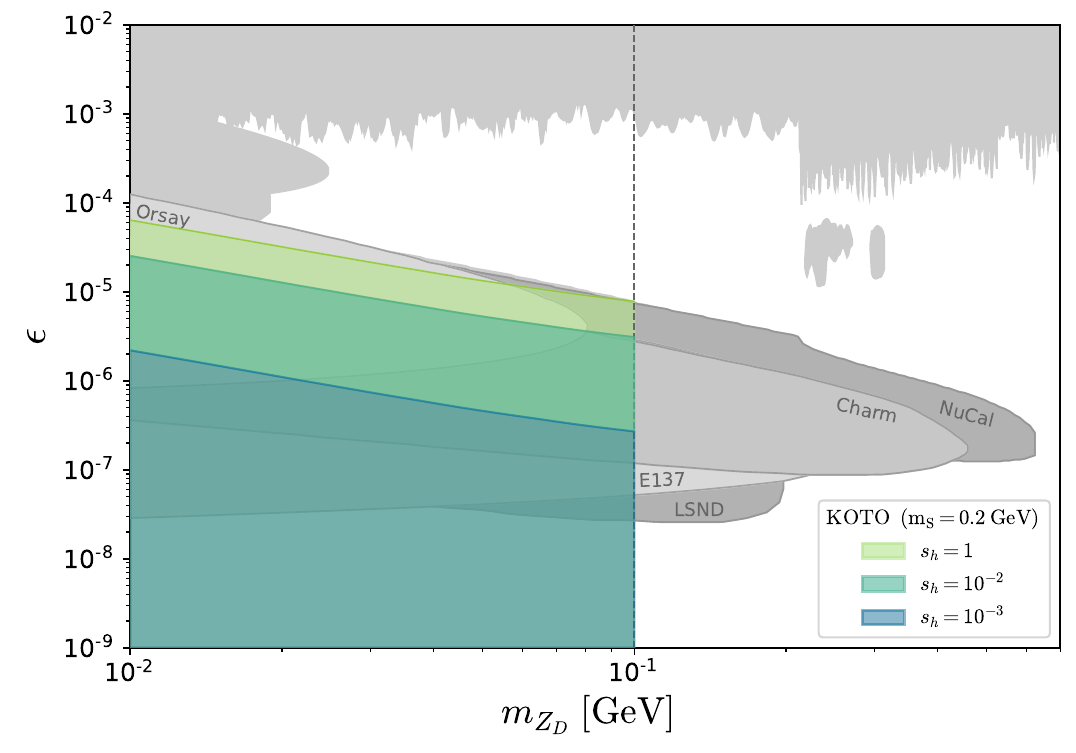}
\end{center}
\vglue -0.8cm
\caption{\label{fig:KOTOeps} Limits from KOTO translated to the dark photon parameter space of the HAHM, where we have fixed $m_S=0.2$ GeV and colored regions are excluded for $s_h=1$ (light green), $10^{-2}$ (green) and $10^{-3}$ (blue green). The regions extend down to $\epsilon=0$. The other gray bounds are dark photon limits coming from the following experiments: APEX~\cite{APEX:2011dww}, A1~\cite{A1:2011yso,Merkel:2014avp}, PS191~\cite{Bernardi:1985ny}, NuCal~\cite{Blumlein:1990ay,Blumlein:1991xh}, CHARM~\cite{CHARM:1985anb}, E137~\cite{Bjorken:2009mm,Andreas:2012mt}, Orsay~\cite{Andreas:2012mt}, BaBar~\cite{BaBar:2014zli}, KLOE~\cite{Anastasi:2015qla,Anastasi:2018azp,KLOE-2:2012lii}, BESIII~\cite{BESIII:2017fwv}, LHCb~\cite{LHCb:2017trq,LHCb:2019vmc}, NA48~\cite{NA482:2015wmo} and LSND~\cite{LSND:1997vqj,Bauer:2018onh}.}
\end{figure}

We show in figure~\ref{fig:KOTOlim} these limits for some choices of the kinetic-mixing parameter $\epsilon$ and the dark photon mass $m_{Z_D}=0.01$ GeV. We see that, in general, the bounds depend strongly on the dark photon parameters, in particular, KOTO looses sensitivity as $\epsilon$ grows and gains 
sensitivity as the kinetic-mixing diminishes until saturating around $\epsilon\sim 10^{-6}$ (for $\epsilon \lesssim 10^{-6}$ all $Z_D$'s will decay outside KOTO). This behavior owes to the fact that the KOTO experiment is prone to measure/constrain an invisible signal; if the dark photons decay inside the detector, the event is vetoed and does not contribute to the signal. As such, for any point in the $m_S\times s_h$ plane that is excluded, we can also exclude all points in the $m_{Z_D}\times \epsilon$ plane for which the dark photons have a larger life-time. This point is made more explicit in figure~\ref{fig:KOTOeps}, where we plot the excluded region in the dark photon parameter space for some values of $s_h$ and $m_S=0.2$ GeV. Contrary to what one would naively expect, the experiment excludes all the values of $\epsilon$ below a certain maximal value which is a function of $m_Z{_D}$\footnote{For values of $\epsilon$ between $10^{-10}$ and $10^{-17}$ we also have bounds coming from Big-Bang Nucleosynthesis and Cosmic-Microwave-Background \cite{Fradette:2014sza,Forestell:2018txr}.}. However, these upper bounds are conditional upper bounds: they depend on $s_h$.
In figure~\ref{fig:KOTOlim} we also present an estimate of the future sensitivity of KOTO to the model for values of the kinetic-mixing smaller than $\sim 10^{-6}$, considering that the experiment will be able to measure the SM branching ratio ${\rm BR}(K_L\to \pi^0 \bar \nu \nu)$. It is thus clear that KOTO is already able to constraint the dark photon HAHM parameter space 
in a non-trivial way for $m_{Z_D}\lesssim 0.1$ GeV,  and that it will in the future be able to put novel constraints on the dark Higgs parameters as well for $m_S \lesssim 0.3$ GeV.

%
%

\subsection{LHCb}\label{sec:lhcb}
The LHCb experiment at the Large Hadron Collider (LHC) at CERN was conceived to perform precision measurements of CP violation and rare $B$ meson decays. They have studied $B^{\pm}$ production in $pp$ collisions at the center of mass energies of 7 TeV and 13 TeV, reporting the production cross-sections $\sigma(pp \to B^{\pm}X,\sqrt{s}=7\;  \rm TeV)=(43.0 \pm 3.0)\,  \mu$b and $\sigma(pp \to B^{\pm}X,\sqrt{s}=13\;  \rm TeV)=(86.6 \pm 6.4)\,  \mu$b summed over both charges and integrated over the transverse momentum range $0<p_T<40$ GeV and rapidity range $2.0 < \eta < 4.5$~\cite{LHCb:2017vec}. These measurements which correspond, respectively, to 1.0 fb$^{-1}$ and 0.3 fb$^{-1}$ imply that about a few $10^{10}$  pairs of $B^\pm$ were produced at each luminosity. 
This makes the  LHCb detector ideal to look for $B \to K S$ decays\footnote{We do not consider decays of neutral states, namely $B^0/\bar{B}^0$, because the identification and reconstruction of $K_{L,S}$ is experimentally much more challenging. We estimate that these modes, however, would only increase by a few percent the sensitivity we present here.} in the mass range $0.6 \lesssim m_S/{\rm GeV}\lesssim 5$. This is even more so if we think about the future  upgrades that predict that the LHCb experiment will have accumulated a data sample corresponding to a minimum integrated luminosity of 300 fb$^{-1}$~\cite{LHCb:2018roe} by the start of the next decade.

%
\begin{figure}[t]
\begin{center}
\includegraphics[width=0.8\textwidth]{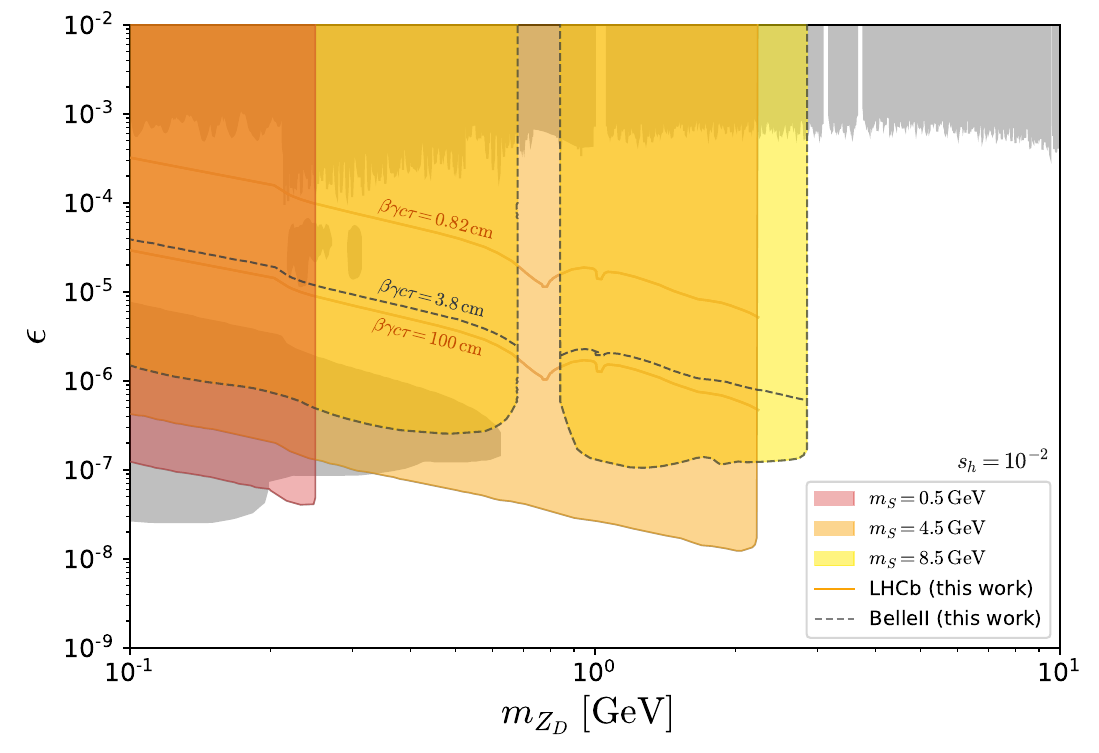}
\end{center}
\vglue -0.8cm
\caption{\label{fig:BepsMzd} Expected sensitivity in the $m_{Z_D}\times\epsilon$ parameter space for the LHCb (solid) and Belle II (dashed) searches. The colors represent different dark photon mass choices, as indicated by the labels. The dashed black line crossing Belle II bound indicates where the dark photon decay length reaches $3.8\, {\rm cm}$, which is the distance at which it enters the vertex detector, such that the signal can be recognized as displaced. Similarly, we indicate with the upper solid orange line the distance of the LHCb vertex discriminator. The lower solid line denotes the distance at which the dark photons exit the vertex detector of LHCb. Gray bounds represent the same dark photon exclusion limits as described in figure \ref{fig:KOTOeps}.}
\end{figure}

Taking the $B^\pm$ mesons to decay into kaons and a dark Higgs, the signature we look for in  LHCb is therefore $B^\pm\to K+4\ell$, where $K$ includes several different kaon states (see below). Opposed to the case of KOTO, the 4 leptons in the final state must be reconstructed in order for the corresponding branching ratio to be measured, thus characterizing a visible signal. As discussed in section~\ref{sec:limHad}, the formula for the number of events in this case is given by eq.~\eqref{eq:Nevts}, with the relevant branching ratio being
\begin{equation}\label{eq:BtoKs}
    {\rm BR}(B^\pm \to K S) =s_h^2\, \frac{\vert g_{sb}\vert^2}{\Gamma_{B}}\frac{2\, p_S^{BK}}{m_{B}}\frac{\vert \langle K\vert \bar{s}_L b_R\vert B^{\pm}\rangle\vert^2}{16 \pi\,  m_{B}}\, ,
\end{equation}
where $p_S^{BK}=\lambda^{1/2}(m_B^2,m_K^2,m_S^2)/(2 \, m_B)$, 
$\Gamma_{B}$ is the $B^\pm$ total decay width, $m_B$ is the $B^\pm$ mass, $m_K$ is the mass of the appropriate kaon state, $g_{sb}$ is the contribution from the electroweak $sbW$-loop diagram 
\begin{equation}
    g_{sb} = \frac{m_b}{v} \frac{3\sqrt{2}\, G_F}{16 \pi^2} \sum_{i=uct} V^*_{is} \, m_i^2\,  V_{ib},
\end{equation}
and the matrix elements $\langle K\vert \bar{s}_L b_R\vert B^{\pm}\rangle$ depend on the parity and spin of the final kaon. We summarize all the expressions for them in appendix~\ref{app:BtoK}. Then, in eq.~\eqref{eq:Nevts} the geometric acceptance is given approximately by $2 \lesssim\eta\lesssim 5$ and the baseline distances taken for the detector are $l_\text{min}=0$ and $l_\text{max}=20$ m \cite{LHCb:2008vvz}. Finally, the efficiency is the combination of the efficiencies for particle identification and reconstruction, which are respectively given by 0.9/0.97/0.95 and 0.96 for electrons/muons/kaons \cite{LHCb}. The values for these efficiencies hold if the respective particles fall inside the vertex locator detector near the interacting point (up to 1 m along the beam axis). If the dark photons are sufficiently long-lived such that they decay outside the vertex locator, we expect that the reconstruction efficiency diminishes.

\begin{figure}[t]
\begin{center}
\includegraphics[width=0.8\textwidth]{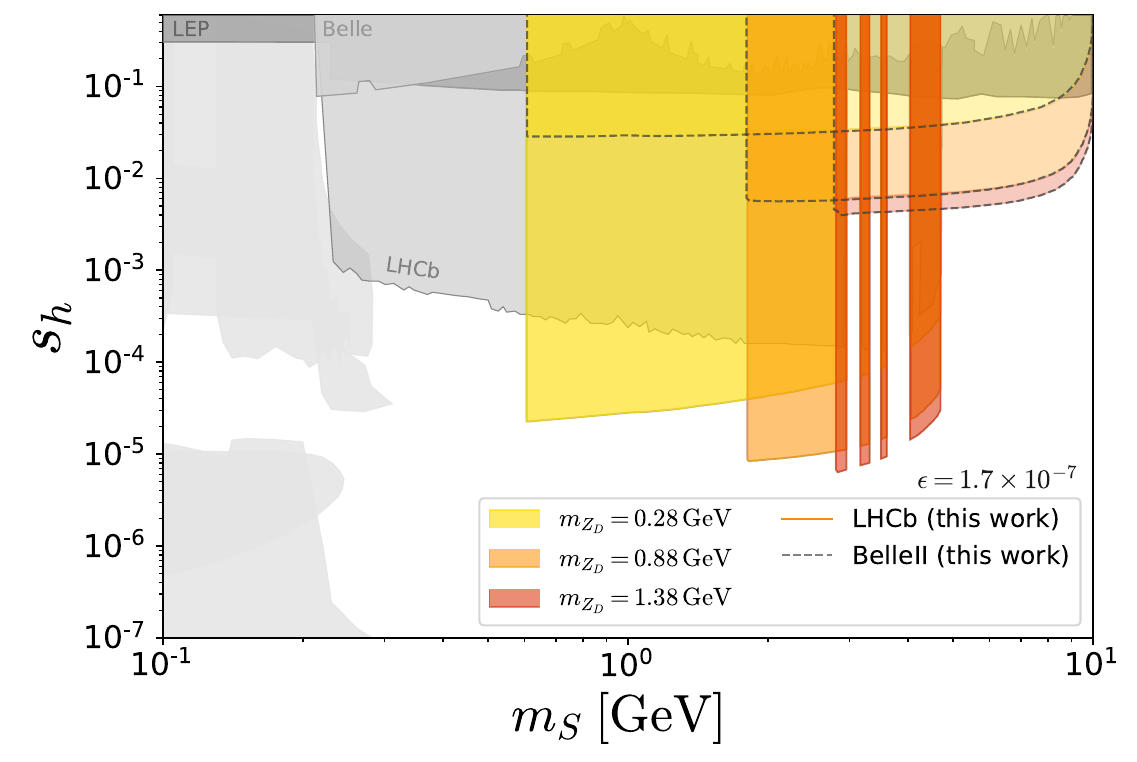}
\vspace{4pt}
\includegraphics[width=0.8\textwidth]{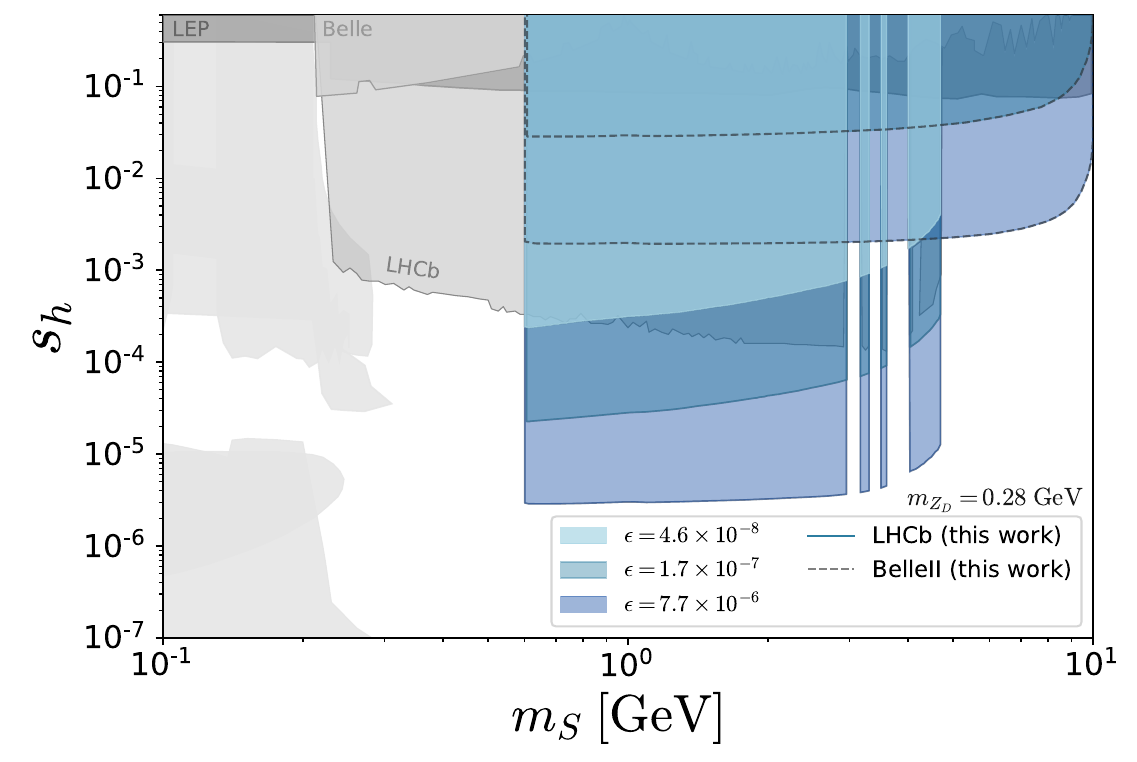}
\end{center}
\vglue -0.5cm
\caption{\label{fig:BsinTeps} In the upper panel we show the expected bounds in the $m_S\times s_h $ plane for the LHCb search (solid) and the future sensitivity projection for the Belle II search (dashed). The colors denote different dark photon mass choices, as indicated in the labels. We fixed the kinetic mixing parameter to $\epsilon = 1.6 \times 10^{-7}$. The lower panel shows the same projection, but for a fixed dark photon mass of  $m_{Z_D} = 0.28 \GeV$ and  varying  kinetic mixing parameter values according to the labels. In both figures, the vetoed regions represent the meson resonances that we have  considered as irreducible backgrounds and the gray bounds are similar to the ones of figure~\ref{fig:KOTOlim} with the addition of the limits from Belle~\cite{Belle:2021rcl} and LEP~\cite{L3:1996ome}.}
\end{figure}

In order to compute exclusion regions in the parameter space, we need to have control over the background coming from the SM. Though the process $B^\pm\to K+4\ell$ can take place in the SM, just as $B^\pm\to K+2\ell$ does, the former has not
yet been experimentally observed and we can thus only rely on theoretical estimates for its branching ratio. The most naive estimate\footnote{A recent calculation for kaon decaying to 4 leptons reads ${\rm BR}(K^+\to \pi^++4\ell)\sim \order{10^{-11}}$ \cite{Husek:2022vul}.} is given by ${\rm BR}(B^\pm\to K+4\ell)_\text{SM}\ll {\rm BR}(B^\pm\to K+2\ell)_\text{SM}\simeq 10^{-6}$, while the same branching ratio  is in the HAHM case roughly ${\rm BR}(B^\pm\to K+4\ell)_\text{HAHM}\simeq 0.5\times s_h^2$ for a prompt signal and $m_S=0$. Whence, we see that if $s_h\gtrsim 10^{-3}$, the signal of the HAHM is much larger than the SM one and we can thus neglect the SM contribution to the background. If, however, the process in the SM and in the model have similar branching ratios, it may be quite difficult to distinguish them if the decay is prompt. In this case, to be able to ignore the SM background we can rely on the long-lived nature of the dark photons and impose the signal to be exclusively displaced, \textit{i.e.} that the decays of the dark photons occur some distance away from the interaction point. This requirement puts a lower bound on the decay length $d=\beta \gamma c \tau$ of the dark photons that can be probed, whose precise value depends on the spatial resolution of each experiment. In the case of LHCb, this value is  approximately 0.8 cm \cite{LHCb:2008vvz}. Note, however, that this requirement could be relaxed if one used the distribution of the reconstructed mass of the pair of leptons with the same flavor, to discriminate between signal and background. This distribution would have a peak at the $m_{Z_D}$ mass in the case of the HAHM signal that could be used to mitigate the SM background. We will refrain from including this here, but this could potentially improve the final sensitivity.

\begin{figure}[t]
\begin{center}
\includegraphics[width=0.7\textwidth]{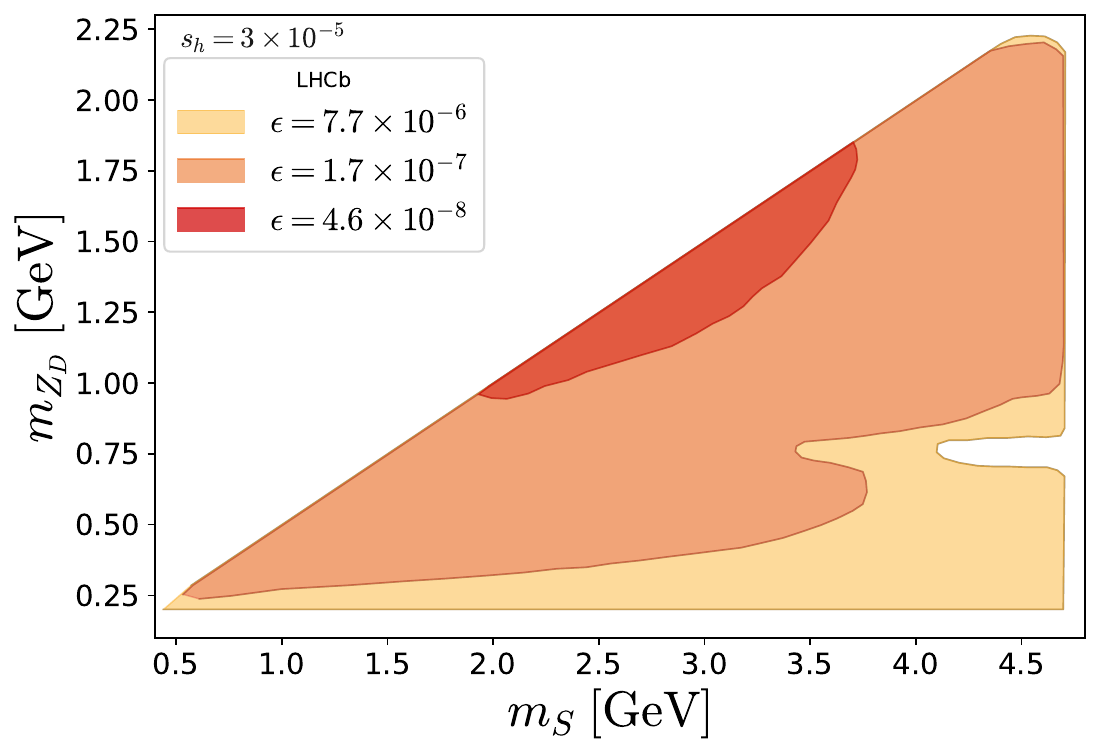}
\vspace{4pt}
\includegraphics[width=0.7\textwidth]{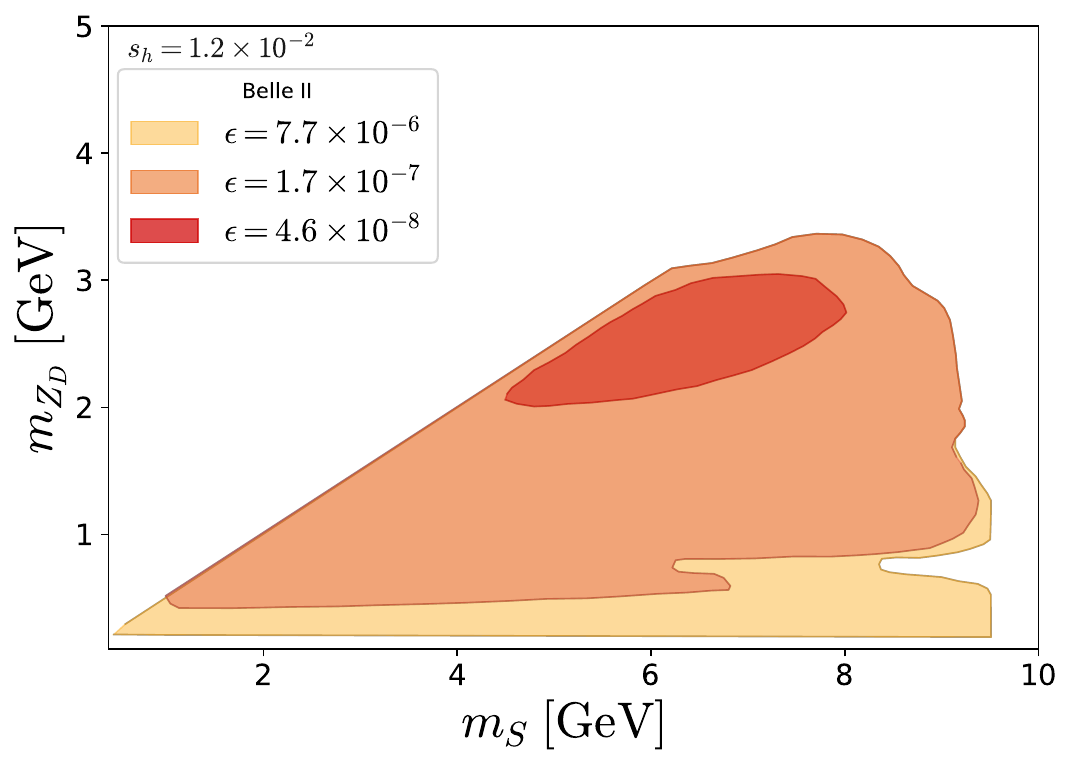}
\end{center}
\vglue -0.5cm
\caption{\label{fig:BoundMass} Regions on the $m_S\times m_{Z_D}$ parameter space for the LHCb (upper panel) and Belle II (lower panel) searches that satisfy $N_{\rm evts}\geq 3$. The colors denote different choices for the kinetic mixing value, as indicated in the labels. The sine of the mixing angle was fixed to $s_h =3 \times 10^{-5}$ ($s_h =1.2 \times 10^{-2}$) in the upper (lower) panel.}
\end{figure}

\begin{figure}[t]
\begin{center}
\includegraphics[width=0.8\textwidth]{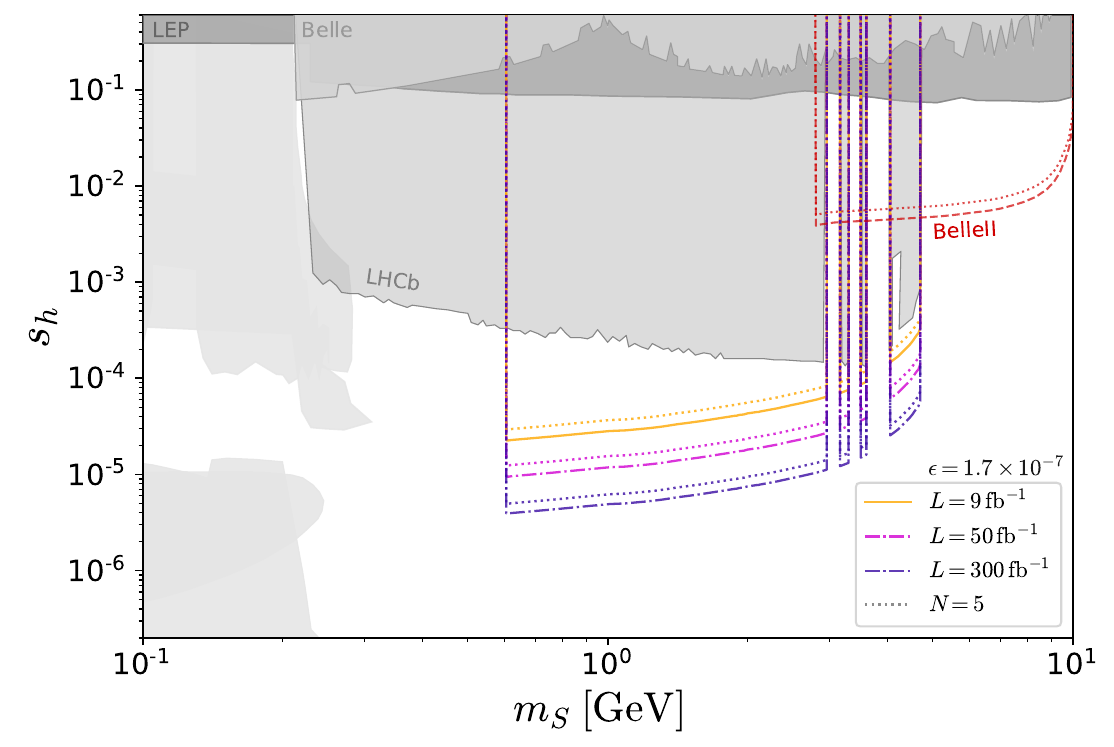}
\end{center}
\caption{\label{fig:BsinTepsMzd} Projected sensitivities on the dark Higgs parameter space for the LHCb (solid) and Belle II (dashed) searches considering the current LHCb integrated luminosity (orange), and the future reach perspectives of $L= 50 \, {\rm fb}^{-1}$ (pink) in Run 3-4 and $L= 300 \, {\rm fb}^{-1}$ (purple) in Run 5-6~\cite{LHCb:2018roe}. The dotted lines are the corresponding $N=5$ events curves that represent the discovery sensitivities. We fixed the dark photon mass to $m_{Z_D}= 0.28 \GeV$ for the LHCb curves and to $m_{Z_D}= 1.38 \GeV$ in the Belle II curves. The kinetic mixing parameter was fixed to $\epsilon= 1.7 \times 10^{-7}$. The remaining gray excluded regions are similar to the ones in figure~\ref{fig:BsinTeps}.} 
\end{figure}

To compute the number of events from eq.~\eqref{eq:Nevts} we have simulated a $B^\pm$ meson flux with \verb!Pythia8! \cite{Sjostrand:2007gs} using the heavy quark mode. The implementation of the detector geometry and the computation of the decay probability was done through \verb!MadDump!~using the UFO model of ref. \cite{Curtin:2014cca}.

Our results are shown in figures~\ref{fig:BepsMzd} and \ref{fig:BsinTeps}. In figure~\ref{fig:BepsMzd} the solid curves represent the exclusion regions in the dark photon parameter space at 95\% C.L. for LHCb, that corresponds to 3 events. We fix $s_h=10^{-2}$ to avoid the SM background in the prompt region, and we indicate the solid line corresponding to $\beta\gamma c\tau=0.82$ cm, below which the signal is displaced. From the plot it is clear that we can achieve a very large sensitivity in $\epsilon$, which is due to the fact that the production depends only on $s_h^2$ and not on $\epsilon$. We also observe a strong dependence on the mass of the dark Higgs, that in particular sets a maximum value for the dark photon mass. In figure~\ref{fig:BsinTeps} instead we show the same in the dark Higgs parameter space, fixing the value of $\epsilon$ ($m_{Z_D}$) on the upper (lower) panel, while the colors denote different choices of $m_{Z_D}$ ($\epsilon$). All values of $\epsilon$ are chosen in order to give displaced signals. We see that we lose sensitivity as we raise the mass of the dark photon, due to the kinematics of $S\to Z_DZ_D$, and as we lower the kinetic-mixing, since then the $Z_D$'s start to decay outside the detector. It is important to note that although we indicate in gray previous dark Higgs searches, they should be viewed with caution as they were not derived for the HAHM model where 
the dark Higgs  decays in a completely different way. We can see this point explicitly for the case of the LHCb, in which the previous search in gray and our bounds cover different regions of the parameter space.

In figure~\ref{fig:BoundMass} we show instead the contour regions for a given number of events in the $m_{S}\times m_{Z_D}$ plane while fixing the couplings, here each colored region satisfy $N_{\rm evts}\geq 3$ for distinct values of $\epsilon$. From such a plot we can better understand the kinematics of this process and its interplay with different experiments. On the upper panel we show the contours for LHCb and bring attention to three features: \textit{i)} the curves vary smoothly with $m_S$, as ${\rm BR}(S\to Z_DZ_D)\simeq 1$ and $P_{\rm dec}$ is independent of the dark Higgs width; \textit{ii)} the dependence on $m_{Z_D}$ is very sizable, due to $N_{\rm evts}\propto {\rm BR}(Z_D\to \ell^-\ell^+)^2$; \textit{iii)} the number of events grow with the masses and are the largest near the threshold $m_S\to 2 \, m_{Z_D}$. This latter point follows from the fact that as $m_{S}\to 2\, m_{Z_D}$, the dark photons are produced more collimated and thus tend more frequently to fall inside the geometric acceptance of LHCb.

Finally, in figure~\ref{fig:BsinTepsMzd} we show by the dotted lines the projected discovery sensitivity (5 events) in the plane $m_S \times s_h$ for the LHCb experiment assuming $m_{Z_D}= 0.28$ GeV and $\epsilon = 1.7 \times 10^{-7}$ and three different integrated luminosities $L= 9$ (orange), 50 (pink) and 300 (purple) fb$^{-1}$~\cite{LHCb:2018roe}.

\subsection{Belle II}

The Belle II experiment~\cite{Belle-II:2010dht} is operating at the asymmetric SuperKEKB $e^+e^-$ (4 GeV/7 GeV) collider aiming to collect about a billion $\Upsilon$(4S) events in about ten years of data taking. The idea is to work in the intensity frontier to try to observe signatures of BSM particles and processes by means of measuring rare flavor physics reactions at lower energies with very high precision and at the same time improve the measurements of the SM parameters~\cite{Belle-II:2018jsg}. Although the range of beam energies covers from the $\Upsilon$(1S) up to the $\Upsilon$(6S) resonance, the main contributions for our signal comes from $\Upsilon$(1S, 2S, 3S) as they have larger branching ratios to leptons. We will assume here Belle II will take data equivalent to 40 times Belle running luminosity in each of these modes according to table 2 of ref.~\cite{Belle-II:2018jsg}, i.e. 240 fb$^{-1}$, 10$^{3}$ fb$^{-1}$ and 120 fb$^{-1}$, respectively, for $\Upsilon$(1S), $\Upsilon$(2S) and $\Upsilon$(3S), corresponding to $4\times 10^{9}$, $6.3 \times 10^9$ and $4.8 \times 10^{8}$ events.

The signal here is $\Upsilon \to \gamma S$ and the corresponding branching ratio can be obtained in terms of BR$(\Upsilon \to e^+ e^-)$ as

\begin{equation}
    {\rm BR}(\Upsilon \to \gamma S) =  
    \frac{s^2_h m_b^2 G_F}{\sqrt{2}\pi \alpha} \, \left[1 - \frac{m_S^2}{m_{\Upsilon}^2} \right] \, {\rm BR}(\Upsilon \to e^+ e^-)\, ,
    \label{eq:UpsilontogammaS}
\end{equation}
where $\alpha= e^2/4\pi$ is the fine structure constant and $m_\Upsilon = 9460, 10023$ and 10355 MeV for $\Upsilon$(1S), $\Upsilon$(2S) and $\Upsilon$(3S), respectively. Both $\Upsilon {\rm(1S,2S)} \to e^+ e^-$ have been experimentally observed with BR$(\Upsilon({\rm {1S}}) \to e^+ e^-)=0.0238$ and  BR$(\Upsilon({\rm{2S}}) \to e^+ e^-)=0.0191$, while $\Upsilon ({\rm 3S}) \to e^+ e^-$ has been seen and BR$(\Upsilon({\rm 3S}) \to e^+ e^-)$ is estimated to be 0.021 according to the Review of Particle Physics~\cite{Workman:2022ynf}. From this we  expect a branching ration BR$(\Upsilon \to e^+ e^-) \sim 4\times 10^{-4} s_h^2$ for $m_S \to 0$. We also expect a non-vanishing SM background in this case. The four lepton electromagnetic decay of the quarkonia $\Upsilon \to \ell^+ \ell^-\ell'^+ \ell'^-$ has not yet been observed, but they are predicted to have a branching ratio of about $10^{-5}$~\cite{Chen:2020bju}, from which we estimate BR($\Upsilon \to \ell^+ \ell^-\ell'^+ \ell'^- \gamma) \sim 10^{-7}$ for the SM. To avoid this background we can use the same strategies as in section~\ref{sec:lhcb}. If we search for displaced signals, we will select signal events when the dark photon decays after the innermost layer of the silicon vertex detector at 3.8 cm. If the signal is prompt, we can either impose that the HAHM signal is much larger than the SM contribution, which in this case holds for $s_h\gtrsim 10^{-2}$, or by studying the kinematical distributions of the leptons \footnote{In this particular case of vector meson decays, the invariant mass distribution of lepton pairs in the SM would present a slight linear growth in the region $3<m_{Z_D}/{\rm GeV} <9$ (see figure 5 of ref.~\cite{Chen:2020bju}).}.

For the computation of the decay probability $P_{\rm dec}$ in eq.~\eqref{eq:Nevts}, we implemented inside {\verb!MadDump!} the Belle II spectrometer volume as a 738 cm long cylinder with a radius of 348 cm. The interaction point is dislocated 45 cm from the center of the detector in the $z$ (beam) direction due to the beam asymmetry. Furthermore, the geometric acceptance of the experiment is $17^\circ <\theta<150^\circ$, where $\theta$ is the polar angle with respect to the displaced interaction point~\cite{Belle-II:2010dht}. For the detection efficiency, we considered a factor of $0.95$ for each lepton identification~\cite{Belle-II:2018jsg}, which accounts to a overall factor of $\varepsilon = (0.95)^4$.

In figure~\ref{fig:BepsMzd}  we show the predicted sensitivity for the signal at  Belle II in the plane $m_{Z_D} \times \epsilon$ for $m_S=8.5$ GeV and $s_h=10^{-2}$. There we also show a dashed line for $\beta \gamma c \tau = 3.8$ cm, marking the dark photon decay length $d$ corresponding to the entrance of the vertex detector. Beyond this point the signal can be deemed to be displaced at Belle II. Clearly here we can profit from the relatively  large quarkonia masses to access $m_{Z_D}$ up to $\sim 5$ GeV but the signal dies for $m_S \gtrsim 2\, m_{Z_D}$ as in this case the $Z_{D}$'s are produced collimated and so can escape detection. Similarly, in figure~\ref{fig:BsinTeps} we show by a dashed line the sensitive region in the plane $m_S \times s_h$ for Belle II and $\epsilon = 1.7 \times 10^{-7}$ (upper panel) and $m_{Z_D}= 0.28$ GeV (lower panel). Although the sensitivity to $s_h$ is about three orders of magnitude less than that of LHCb, it extends to $m_S \sim 10$ GeV in a region not covered by any other experiment.

As before, we show in the lower panel of figure~\ref{fig:BoundMass} the $N_{\rm evts}\geq 3$ contour regions in the $m_S\times m_{Z_D}$ plane for Belle II. We notice that the contours have similar properties to the ones of LHCb in the left panel, except for the loss of sensitivity in the large mass region as stressed previously. We can also see in figure~\ref{fig:BsinTepsMzd}, as an illustration, the expected discovery sensitivity (5 events) of Belle II to the HAHM dark Higgs in the plane $m_S \times s_h$, for $m_{Z_D}=1.38$~GeV  and $\epsilon = 1.7 \times 10^{-7}$  (red dotted line).

\section{Higgs Invisible Decay}
\label{sec:HiggsInv}

In the HAHM  the SM Higgs-like scalar $h$ can decay to dark particles, that can potentially contribute to its invisible width as long as they evade detection. The search for invisible decays of the SM Higgs at the LHC resulted in the current limit BR$^{\rm exp}(h \to \rm invisible)<0.19$~\cite{CMS:2018yfx}. On the other hand, its total decay width $\Gamma^{\rm exp}_{\rm total}=3.2\pm^{2.8}_{2.2}$ MeV has been experimentally obtained by CMS using the on-shell and off-shell production in four-lepton final states under the assumption of a coupling structure similar to the SM one~\cite{CMS:2019ekd}. These experimental results allow  us to use our  predictions for the invisible branching ratio of the HAHM at CMS to constrain a new region of the model parameter space.

Analogous to eq.~\eqref{eq:KOTO_BR_eff} we have used in the case of the KOTO experiment, we define an effective branching ratio of $h$ as
\be
{\rm BR}_{\rm eff}^{\rm inv} \equiv {\rm BR}(h \to {\rm invisible})\vert_{\rm HAHM} = \frac{1}{\Gamma^{\rm exp}_{\rm total}} \sum_{{\rm channel}~i} P^{\rm out}_i\, \Gamma(h\to i)\, ,\label{eq:h_BRinv}
\ee
where $i$ is a final state containing only dark particles, $P^{\rm out}_i$ is the corresponding probability for them to decay outside the detector and the sum is performed over all possible channels. Here we maintain our original assumption that the dark Higgs decays instantaneously inside the detector, such that only dark photons can be long-lived. Restricting ourselves to first order in $s_h$, on-shell intermediate states only and neglecting processes with extra suppression factors of $\epsilon$ or $1/v^2$, we find 4 possible channels: $h\to Z_DZ_D$, $h \to SS \to 4Z_D$, $h\to SZ_DZ_D\to 4Z_D$ and $h\to 3S \to 6Z_D$. The respective widths in the limit $m_{Z_D},m_S\ll m_h$ are
\be\label{eq:hZDZD_app}
\Gamma(h\to Z_DZ_D)\simeq \Gamma(h\to SS)\simeq \left(\frac{g_D s_h}{m_{Z_D}}\right)^2\frac{m_h^3}{32\pi},
\ee
and 
\be\label{eq:sZDZD_app}
\Gamma(h\to S Z_DZ_D)\simeq \frac{(g_D^2s_h)^2}{3(64)^2\pi^3}\frac{m_h^5}{ m_{Z_D}^4}\left(1-12 \frac{m_{Z_D}^2}{m_h^2} + 108\frac{m_{Z_D}^4}{m_h^4}\right),
\ee
independent of $m_S$, the last expression being valid for $m_S \lesssim 10$ GeV. We do not 
show here the width for the case of the decay into 3 pairs of $Z_{D}$'s through $h \to SSS$ as it can be safely ignored. See appendix~\ref{app:HAHM} for the complete expressions for all these widths as well as a visual illustration of their relevance given in figure~\ref{fig:WidthHinv}. Since we assume $S$ to be on-shell, the subsequent decay $S\to Z_DZ_D$ in eqs.~\eqref{eq:hZDZD_app} and \eqref{eq:sZDZD_app} is obtained by multiplying the widths by the appropriate power of ${\rm BR}(S\to Z_DZ_D)$, which is very close to $1$ since we assume $g_D$ satisfies eq.~\eqref{eq:condition_sZDZD}.

Before describing how we have simulated these decays at CMS in order to calculate $P_i^{\rm out}$, let us comment on the size of each of these contributions to the $h$ invisible decay width. We see from eq. \eqref{eq:hZDZD_app} that the two main contributions to the total invisible decay width are $\Gamma(h\to Z_DZ_D)\simeq \Gamma(h\to SS)$, which are practically independent of $m_S$ and will impose a constraint on $(g_D s_h)/m_{Z_D}$. The channel $h \to S Z_D Z_D$, is proportional to $g_D^4\,  m_{Z_D}^{-4}$, instead of to $g_D^2\, m_{Z_D}^{-2}$ as the two aforementioned major channels. As a consequence, the latter channel will be more important at lower values of $m_{Z_D}$ as long as $g_D\gtrsim 10^{-1}$. The last mode, $h \to SSS$, is at least two orders of magnitude smaller than the previous channels and can be neglected (see appendix~\ref{app:HAHM} for the explicit expression and comparison with  other decay widths).

%
\begin{figure}[t!]
\begin{center}
\includegraphics[width=0.7\textwidth]{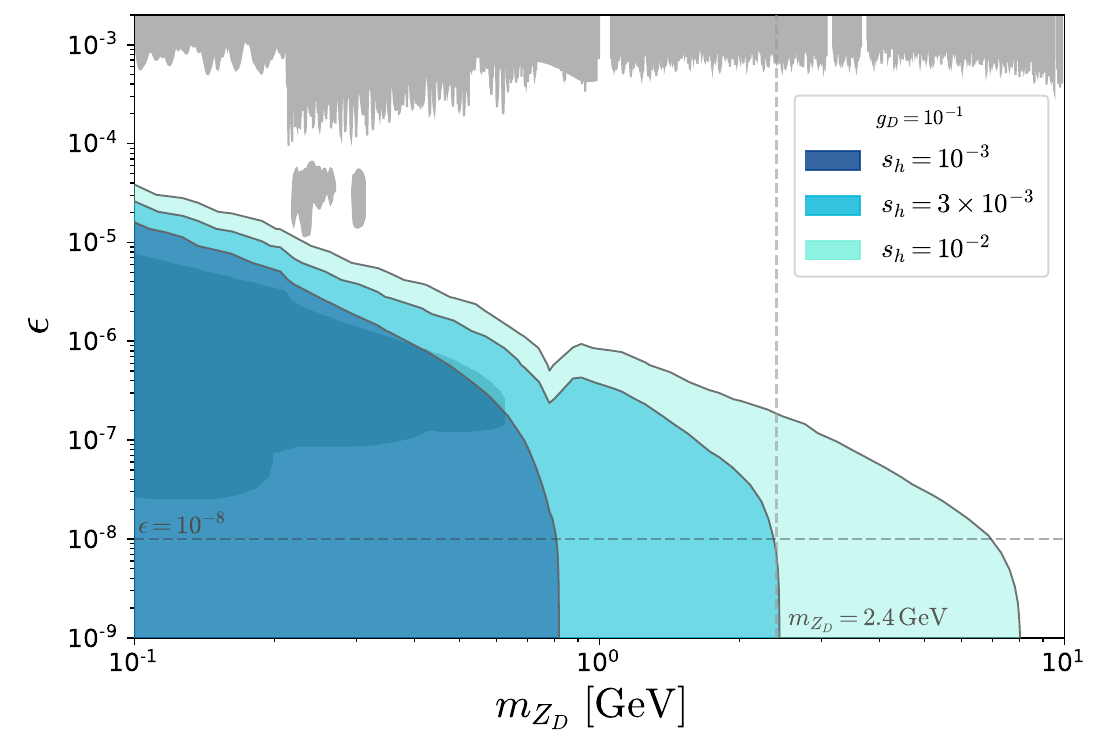}
\vspace{4pt}
\includegraphics[width=0.7\textwidth]{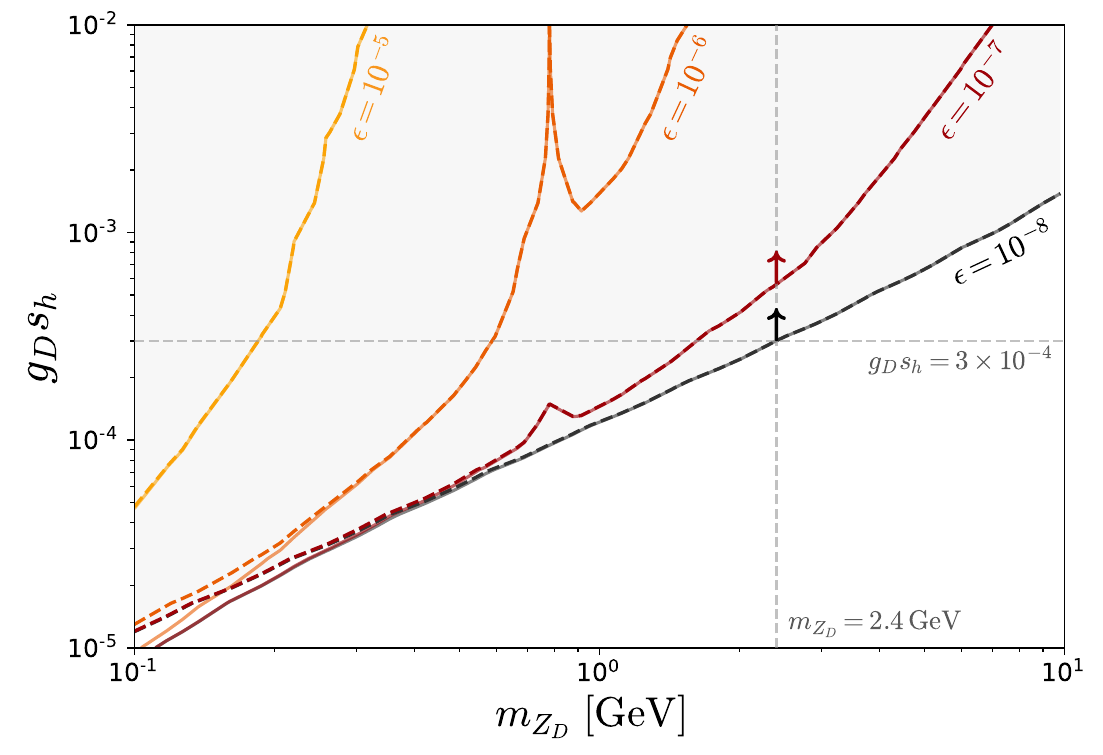}
\end{center}
\vglue -0.5cm
\caption{\label{fig:HInv1} In the upper panel we show the bound on $m_{Z_D}\times \epsilon$ obtained by using the CMS limit ${\rm BR}_{\rm eff}^{\rm inv}<0.19$~\cite{CMS:2018yfx} for $s_h= 10^{-3}$ (blue), $3\times 10^{-3}$ (cyan) and $10^{-2}$ (light blue) and dark gauge coupling fixed to $g_D= 10^{-1}$. In the lower panel we show the same, but in the plane $m_{Z_D}\times g_D s_h $ and for $\epsilon = 10^{-5}$ (yellow), $10^{-6}$ (orange), $10^{-7}$ (red) and $10^{-8}$ (black). The dependence on the dark Higgs mass $m_S$ can be neglected in the mass range considered here. In the upper (lower) panel the curves are limited from above (below) by the condition that the dark photons must decay outside the CMS detector, but are unlimited from below (above). For a fixed value of $m_{Z_D}$ we can read from the lower plot the exclusion limit on $g_D s_h$ for different choices of kinetic mixing $\epsilon$, as indicated by the arow for $m_{Z_D} = 2.4$ GeV (vertical dashed grey line).  }
\end{figure}

The decay probabilities in eq.~\eqref{eq:h_BRinv} were computed by simulating the SM-like $h$ boson in the vector boson fusion (VBF) production mode using \verb!MadGraph5!~\cite{Alwall:2014hca}, assuming a center of mass energy of $\sqrt{s}=$ 13 TeV. According to ref. \cite{CMS:2018yfx}, the VBF production channel is not the only relevant one to the determination of ${\rm BR}^{\rm exp}(h\to {\rm invisible})$, which includes gluon fusion and associated production as well. However, since VBF is the channel that most contribute to the bound, we use it as a benchmark to make our predictions. We have checked that the kinematical distributions of the simulated $h$ are in good agreement with refs.~\cite{CMS:2018yfx,LHCHiggsCrossSectionWorkingGroup:2016ypw}. We then used these distributions to calculate the $Z_D$ decay probabilities for each invisible decay channel by implementing the CMS detector geometry \cite{CMS:2008xjf} in \verb!MadDump!.

In the upper panel of figure~\ref{fig:HInv1} we show the exclusion region in the $m_{Z_D}\times \epsilon$ plane for three different scalar mixing values $s_h= 10^{-3}$ (blue), $3\times 10^{-3}$ (cyan) and $10^{-2}$ (light blue) and for $g_D=10^{-1}$. As explained before, these limits do not depend on $m_S$ as long as $m_S \ll m_h$. The upper limit of each curve represents the constraint that the produced dark photons should decay outside the CMS detector. The dip that appears close to $m_{Z_D} = 1$ GeV is a consequence of the appearance of several hadronic resonances in this energy range that affect the dark photon total width (see figure~\ref{fig:DarkWidths}). We can also note that there is no lower bound since the production of dark photons does not depend on $\epsilon$, so by decreasing its value the only effect is to increase the probability of outside decay. Hence, as we have seen for KOTO, the invisible decay sets a minimum allowed value for $\epsilon$ as a function of $m_{Z_D}$ in a complementary way to the limits obtained with LHCb and Belle II. From the figure it is also clear that as we increase the value of $g_D s_h$ we can also exclude larger regions in the parameter space, since we are augmenting the Higgs invisible width.

The lower panel of figure~\ref{fig:HInv1} shows the exclusion region in the $m_{Z_D}\times g_D s_h$ parameter space for four different values of kinetic mixing $\epsilon = 10^{-5}$ (yellow), $10^{-6}$ (orange), $10^{-7}$ (red) and $10^{-8}$ (black). The dashed curves were made including only the two body decay channels $h \to Z_D Z_D$ and $h \to S S$, while the solid curves include also the $h \to S Z_D Z_D$ mode by considering a fixed $g_D= 10^{-1}$. The inclusion of this latter channel is relevant only for $g_D\gtrsim 10^{-1}$ and dark photon masses below 0.2 GeV, and accounts only for a slight deviation compared to the dashed lines. Hence, the mode $h\to SZ_DZ_D$ can be safely neglected and the Higgs invisible width becomes the weighted sum of the 2-body decay channels. Within this approximation and noticing that $P^{\rm out}\leq 1$, the experiment loses sensitivity when
\be \label{eq:limitgdsh}
m_{Z_D} \to g_D s_h \sqrt{(2 m_h^3)/(32 \pi\, \Gamma^{\rm exp}_{\rm total} \, {\rm BR}^{\rm exp}(h\to {\rm invisible}))} \sim 8 \times 10^3 \, g_D s_h.
\ee 
In a similar and more precise way, we can read off from the lower plot the bounds on the couplings while taking into account the effects of the dark photon decay probability. Choosing for example $m_{Z_D} = 2.4$ GeV and $\epsilon = 10^{-8}$, we can exlcude $g_D s_h > 3 \times 10^{-4}$. If we lower the value of $\epsilon$ to $10^{-7}$, this bound is softened to $g_D s_h \gtrsim 5 \times 10^{-4}$.

\section{Discussion and Conclusions}
\label{sec:conc}
We have studied the current limits and future sensitivities of  accelerator experiments at the high intensity and precision frontier to a light realization of the dark sector of the HAHM. This means we have focused here on the scenario where both the dark Higgs $S$ and the dark photon $Z_D$ are light, i.e. have masses below $10$ GeV. The HAHM presents an interesting combination of the scalar and vector renormalizable portals with an additional gauge interaction between the dark Higgs  and the dark photon. This opens up a particularly interesting window that we explore in this paper, namely when the gauge sector dominates the dark Higgs decays. This regime is characterized by $\Gamma_S({\rm total}) \simeq \Gamma(S \to Z_D Z_D)$, which holds for $m_S \geq 2 \, m_{Z_D}$ and for gauge couplings that satisfy $g_D \gg 7 \times 10^{-3} \, s_h$. 

In this scenario we have shown that accelerator experiments have sensitivity to explore the model and access the dark Higgs  ($m_S$ and $s_h$), the dark photon  ($m_{Z_D}$ and $\epsilon$) and the gauge ($g_D$) parameters. We have primarily  studied the possibility of the light dark Higgs being produced by meson (kaons, $B$-mesons and $\Upsilon$'s) decays in facilities like KOTO, LHCb and Belle II. In these experiments, after production,  the dark Higgs  in the assumed scenario can promptly decay  to a pair of dark photons, which can subsequently decay outside or inside the detector volume, depending on the value of the kinetic mixing $\epsilon$. On the one hand, if $Z_D$ is sufficiently long-lived (small enough $\epsilon$), it can decay outside the sensitive volume and be considered an invisible signal. On the other hand, if $Z_D$ is sufficiently short-lived (large enough $\epsilon$), it can decay inside the detector volume. In the latter case we have only explored $Z_D \to \ell^+ \ell^-$ because of the cleaner signature and lower background. This strategy allowed us to decouple the $S$ production mechanism (which depends on the mixing as $s_h^2$) from the decay of the $Z_D$ (which depends on the kinetic mixing as $\epsilon^2$). Each of the examined experiments are prone to scrutinize a different $m_S, m_{Z_D}$ range and so they are complementary.

The rare kaon decay experiment KOTO can potentially provide the best limits for $m_S \lesssim 0.3$ GeV. We have considered KOTO's upper bound on BR$(K_L \to \pi^0 X)$ applied to $S$ and long-lived $Z_D$'s that decay outside the detector. This experiment can currently set limits on the scalar mixing and the kinetic mixing parameters, that are respectively given by $s_h < 7 \times  10^{-4}$ for $m_S \lesssim 0.2$ GeV and $\epsilon \lesssim 7\, (20) \times 10^{-7}$ for $ m_{Z_D}=0.1\,  (0.01)$ GeV. These limit can potentially improve by about an order of magnitude in the future, if KOTO achieves the SM sensitivity to $K_L \to \pi^0 \nu \bar \nu$.

The rare $B$-meson decay experiment LHCb can best explore the model in the range $0.5\lesssim m_S/{\rm GeV}\lesssim 5$. We have evaluated the sensitivity of this experiment to observe the dark Higgs in $B^\pm \to K S \to K + 4\ell$. In this region, current data can set bounds on $s_h \lesssim (10^{-3}-10^{-5})$ depending on $m_{Z_D}$ (except for some values of $m_S$ which correspond to charmonium resonances) and considering that the $Z_D$ decay is displaced (i.e. for $\epsilon \lesssim 10^{-4}-10^{-5}$ depending on $m_{Z_D}$). If the $Z_D$ decay is prompt, limits on $m_{Z_D} \times \epsilon$ can be set if we consider larger values of $s_h$. For instance, if $s_h = 10^{-2}$ one can probe $\epsilon$ down to $10^{-7}$ and even a bit bellow, depending on  $m_{S}$ and $m_{Z_D}$. In the future, if the LHCb experiment collects data corresponding to 50 (300) fb$^{-1}$, these limits can be potentially improved by one (two) order(s) of magnitude.

The asymmetric $e^+e^-$ collider Belle II is positioned to  have the best sensitivity in the range $5\lesssim m_S/{\rm GeV} \lesssim 10$. At Belle II the dark Higgs can  be produced and observed in the  modes $\Upsilon({\rm 1S,2S,3S}) \to \gamma S \to \gamma + 4 \ell$. We have shown that Belle II can probe down to $ s_h \lesssim 3\times( 10^{-3}- 10^{-2})$ depending on $m_{Z_D}$ and on $\epsilon$, as long as the $Z_D$ decay is displaced to avoid SM background. The displacement requirement can perhaps be relaxed by the experiment by using other kinematical distributions, such as the invariant mass of the pair of charged leptons, which is peaked in the case of the signal. This could increase Belle II's  sensitivity to the model to some extent, but hardly by an order of magnitude.

Finally, we have examined the CMS experiment as it can indirectly probe a higher range of dark photon masses, $m_{Z_D} \lesssim 10$ GeV, basically independently of the value of $m_S \ll m_h$. This is also the only experiment that can provide some sensitivity to the gauge coupling $g_D$. Since in the HAHM $h$ is a SM Higgs-like scalar, we were able to use the CMS limit on the BR$^{\rm exp}(h \to {\rm invisible})$ and their measurement of the $h$ total decay width to constrain $g_D s_h$ and $m_{Z_D}$. This is because in the envisage scenario there are several ways $h$ can decay invisibly at CMS, given that $Z_D$ decays outside the detector. As an example, we have shown that for $m_{Z_D}= 2.4$ GeV one can exclude $g_D s_h > 3\times 10^{-4}$ as long as $\epsilon$ is sufficiently small to allow for a long-lived $Z_D$, i.e. $\epsilon \lesssim 10^{-7}$. Note that these CMS bounds could be combined to  Belle II and lead to limits on $g_D$.


\acknowledgments
R.Z.F. were partially supported by Fundação de Amparo à Pesquisa do Estado de São Paulo (FAPESP) under contract 	
2019/04837-9 and Conselho Nacional de Desenvolvimento Científico  e Tecnológico (CNPq). A.L.F was supported by FAPESP under contract 2022/04263-5. G.M.S. was supported by FAPESP under contract 2020/14713-2.

\appendix

\section{Mixing with the SM and decay widths}\label{app:HAHM}

\subsection{Scalar sector mixing}

Let us begin by discussing in more depth the scalar sector of the HAHM model. The relevant interactions are given by the scalar potential in eq.~\eqref{eq:scalar_pot},
\begin{equation*}
     V(H, S) = -\mu^2 |H|^2 + \lambda |H|^4 - \mu_S^2 |S|^2 + \lambda_S |S|^4 + \kappa |H|^2 |S|^2.
\end{equation*}
Assuming that $\mu_S^2,\lambda_S>0$, the dark Higgs develops a vacuum expectation value (vev) $\expval{S}=v_S/\sqrt{2}$ responsible for the spontaneous symmetry breaking of the $U(1)_D$ gauge symmetry, which in turn results in a mass term for the dark photon. Another consequence of the $S$ getting a vev is that there will be mixing between the dark Higgs and the SM Higgs. In order to compute this mass mixing we need to consider the broken phase after both scalars acquire a vev
\begin{equation}
    H \to \begin{pmatrix}
    0 \\
    (v + h_0)/\sqrt{2}
    \end{pmatrix} \,,
\end{equation}
\vspace{5pt}
\begin{equation}
    S \to \frac{(v_S + S_0)}{\sqrt{2}} \,,
\end{equation}
where we already transformed the fields according to unitary gauge. Collecting the quadratic terms of the potential:
\be
V(h_0,S_0)= \frac{1}{2}\left(h_0 ~~ S_0\right)\mathcal{M}_{hs}^2 \begin{pmatrix}h_0\\S_0\end{pmatrix}
\ee
where the mass matrix is given by
\be\label{eq:scalarMmat}
\mathcal{M}_{hs}^2=\mqty(2\, \lambda v^2 &  \kappa \, v \, v_S \\  \kappa \, v \, v_S & 2 \, \lambda_S v_S^2) \,.
\ee
This matrix is diagonalized by the following orthogonal transformation
\be
\mqty(h \\ S)=\mqty(\cos \theta_h & - \sin \theta_h\\ \sin \theta_h & \cos \theta_h )\mqty(h_0 \\ S_0) \,,
\ee
with $h$ and $S$ the physical mass states (where we use the same label convention to the dark Higgs before and after the spontaneous breaking) and $\theta_h$ the scalar mixing angle, defined as
\be
\begin{aligned}
\tan^2{(2 \theta_h )}& =\frac{(2 \, \kappa \, v \, v_S)^2}{(m_h^2-m_S^2)^2-(2 \, \kappa \, v \, v_S)^2}\\
& \simeq \qty(\frac{2 \, \kappa \, v \, v_S}{m_h^2-m_S^2})^2+\order{\kappa^4} \,,
\end{aligned}
\ee
assuming the limit of small $\kappa$. The Higss and dark Higgs squared masses, $m_h^2$ and $m_S^2$, respectively, are the eigenvalues of the mass matrix $\mathcal{M}^2$, given by
\begin{equation}
    m^2_{h,S} = \lambda v^2 + \lambda_S v_S^2 \pm \sqrt{v^4 \lambda^2 +v_S^4 \lambda_S^2 + v^2 v_S^2 (\kappa^2 - 2 \lambda \lambda_S) } \,,
\end{equation}
where we already considered a specific mass hierarchy $m_h > m_S$, which is the relevant one for our study.
Now, in the limit where $\kappa \ll 1$, \ie in the limit of small mixing angles, we have that
\begin{equation}
    \tan 2 \theta_h \simeq  2 \theta_h  \simeq \frac{2 \, \kappa  v  v_S}{m_S^2-m_h^2}
    \,,
\end{equation}
and, hence, $s_h \equiv \sin \theta_h \simeq \theta_h$, such that 
\be\label{eq:scalar_angle}
s_h\simeq \frac{\kappa v v_S}{m_S^2-m_h^2} \,.
\ee
In this limit, the mass eigenvalues become
\begin{equation} \label{eq:hMass}
    m_h^2 = 2 \lambda v^2 + 2 s_h^2 (\lambda v^2 - \lambda_S v_S^2)\,,
\end{equation}
\begin{equation}\label{eq:sMass}
    m_S^2 = 2 \lambda_S v_S^2 - 2 s_h^2 (\lambda v^2 - \lambda_S v_S^2) \,,
\end{equation}
up to order $\kappa^2$. Note that the SM-like Higgs mass would be only corrected from the SM value $m_h^{\mathrm{SM}} = 2 \lambda v^2$ by a small factor proportional to the sine of the mixing angle. The fields will also mix according to
\begin{equation}\label{eq:HSmix}
h_0 = c_h h + s_h S \,, \qquad S_0= c_h S - s_h h \,,
\end{equation}
where $c_h \equiv \cos \theta_h \simeq 1 + \mathcal{O}(\kappa^2)$.

\subsection{Gauge sector mixing}

In the neutral gauge sector, the mixing is realized by the kinetic mixing with the hypercharge gauge boson of eq.~\eqref{eq:KMlag}
\begin{equation*}
    \mathcal{L}_{ZB} = - \frac{1}{4} \hat B_{\mu \nu} \hat B^{\mu \nu}  - \frac{1}{4} \hat Z_{D \mu \nu} \hat Z_D^{\mu \nu} - \frac{\epsilon}{2 c_W} \hat B_{\mu \nu} \hat Z_{D}^{\mu \nu} ,
\end{equation*}
that can be canonically normalized according to the field redefinitions
\bea\label{eq:kinetic_diagonalization}
\hat Z_D^\mu &= \eta \tilde Z_D^\mu, \\
\hat B^\mu = B^\mu& - \frac{\epsilon}{c_W}\eta \tilde Z_D,
\eea
with $\eta=1/\sqrt{1-\epsilon^2/c_W^2}$.
Using eq.~\eqref{eq:kinetic_diagonalization} after the scalars acquire a vev will generate a mass mixing between the gauge bosons. The relevant terms emerge from the kinetic terms of the scalars,
\be
|D_\mu S|^2+|D_\mu H|^2\supset \frac{1}{2}\left(Z^0 ~~\tilde Z_D\right)\mathcal{M}_{ZZ_D}^2\begin{pmatrix}Z^0\\\tilde Z_D\end{pmatrix},
\ee
where we have used eqs.~\eqref{eq:HcovD} and \eqref{eq:ScovD}. In the equation above, $Z^0$ denotes the SM $Z^0$-boson while the mass matrix is
\be
\mathcal{M}_{ZZ_D}^2=m_{Z^0}^2\begin{pmatrix}
1 & \eta t_W\epsilon\\
\eta t_W\epsilon & \eta^2 t_W^2\epsilon^2 + \delta^2
\end{pmatrix},
\ee
with $m_{Z^0}$ the SM $Z^0$-boson mass, $t_W=s_W/c_W$ and $g_Dv_S=m_{Z^0}\delta $. The physical states $Z$ and $Z_D$ are obtained diagonalizing the matrix above via the orthogonal transformation
\be
\begin{pmatrix}Z\\Z_D\end{pmatrix}=\begin{pmatrix}\cos\alpha & \sin\alpha\\ -\sin\alpha & \cos\alpha\end{pmatrix}\begin{pmatrix}Z^0\\\tilde Z_D\end{pmatrix},
\ee
where
\be
\tan 2\alpha = \frac{2 \eta t_W \epsilon}{1-\delta^2-\eta^2t_W^2\epsilon^2}.
\ee
In the scenario where $m_Z \gg m_{Z_D}$, which is the relevant hierarchy we are interested in, and also for suppressed kinetic mixing, we end up with
\begin{equation}
    m_{Z_{D}}^2 \simeq \delta^2 m_{Z^0}^2 (1- \epsilon^2 t_W^2)\, ,
\end{equation}
\begin{equation} \label{eq:massZnew}
    m_{Z}^2 \simeq  m_{Z^0}^2 (1 + \epsilon^2 t_W^2)\, .
\end{equation}
Hence, we can see that the SM-like $Z$-boson mass will suffer a correction proportional to $\epsilon^2$. Similarly we have
\begin{equation}
    Z^0 \simeq Z -  \epsilon t_W  Z_D   \, ,
\end{equation}
\begin{equation}
 \tilde Z_D \simeq  Z_D + \epsilon t_W  Z \, .
\end{equation}

\subsection{Decay widths}

Considering that the masses of the dark particles satisfy $m_{Z_D}\lesssim\order{5~\text{GeV}}$ and $2m_{Z_D}\leq m_{S}\lesssim\order{10~\text{GeV}}$, only a few interactions become relevant. In the neutral gauge sector we have up to first order in the couplings
\bea\label{eq:DP_copling}
\mathcal{L}_\text{NC}&=g'B_\mu J_B^\mu + g W_{3\mu}J^\mu_3\\
&\simeq eA_\mu J^\mu_\text{EM} - \epsilon e Z_{D\mu}J^\mu_\text{EM},
\eea
with $J^\mu_B,~J^\mu_3,~J^\mu_\text{EM}$ the respective currents. Hence, the dark photon couples to the SM via a photon-like current suppressed by an extra factor of $\epsilon$. The scalar sector has many more interactions, but only a few are important for our purposes. These are given by
\bea\label{eq:DH_coupling}
\mathcal{L}_\text{int}\supset g_D m_{Z_D} SZ_{D\mu}Z^\mu_D+s_h \sum_f\frac{m_f}{v}S\bar{f}f,
\eea
where we made use of the relation $m_{Z_D}=g_Dv_S$. The first term above comes from the kinetic term $|D_\mu S|^2$ while the second are the interactions with the fermions inherited from the Higgs. For more details on all other possible interactions, we refer to \cite{Curtin:2014cca}.

With eq.~\eqref{eq:DP_copling} we can compute the partial widths of the dark photon to a pair of fermions:
\be
\Gamma(Z_D\to \bar{f} f)=\frac{(\epsilon e q^f_\text{em})^2N_c }{12\pi}m_{Z_D}\left(1+\frac{2m_f^2}{m_{Z_D}^2}\right)\sqrt{1-\frac{4m_f^2}{m_{Z_D}^2}},
\ee
where $N_c=3(1)$ for quarks (leptons) and $q^f_\text{em}$ is the electromagnetic charge of the fermion. Note that for dark photon masses below $2$ GeV the perturbative calculation to quarks is not valid, since the physical degrees of freedom are hadrons. In this case we use the $R$-ratio \cite{Zyla:2020zbs} to re-scale the width to muons as
\be
\Gamma(Z_D\to \text{hadrons})= R(m_{Z_D}^2)\Gamma(Z_D\to \mu^-\mu^+),
\ee
using that $R(s)=\sigma(e^-e^+\to \text{hadrons})/\sigma(e^-e^+\to \mu^-\mu^+)$ for a center of mass energy $s$. A plot of partial and total widths, together with the respective branching rations is given in figure \ref{fig:DarkWidths}. Note that the branching ratios are independent of $\epsilon$.

The case of the dark Higgs is a bit more subtle. From the mixing in eq.~\eqref{eq:HSmix}, the dark Higgs inherits all Higgs couplings suppressed by a factor of $s_h$. Therefore, the widths to pair of fermions, photons and gluons are given by expressions similar to that of the Higgs. We report here explicitly the decay of $S$ to fermions:
\be\label{eq:sff}
\Gamma(S\to \bar{f}f) = \frac{N_c s_h^2m_f^2}{8\pi v^2}m_S \left(1-\frac{4m_f^2}{m_S^2}\right)^{3/2}.
\ee
As in the case of the dark photon, if $m_S\lesssim 2$ GeV, then the perturbative calculation to quarks and gluons does not hold and we need to perform decays directly to hadrons. However, we do not posses a $R$-ratio for scalars, thence we must rely solely on theoretical computations. To this end, we use the results from ref.~\cite{Winkler:2018qyg}. Finally, given the hierarchy $2m_{Z_D}<m_S$, we have the additional width to a pair of dark photons coming from the first term in eq.~\eqref{eq:DH_coupling}
\be
\Gamma(S \to Z_D Z_D) =  \left(\frac{g_D}{m_{Z_D}}\right)^2 \frac{1}{32 \pi \, m_S} (m_S^4 - 4\, m_S^2 m_{Z_D}^2+ 12 \, m^4_{Z_D})\sqrt{1-  \frac{4m_{Z_D}^2}{m_S^2}}\, .
\ee
We show in figure~\ref{fig:DarkHiggsW} all the partial widths of the dark Higgs, including the ones to hadrons and to two photons.

In section~\ref{sec:HiggsInv} it will be necessary to compute novel decay channels of the Higgs to dark sector particles. The Lagrangian that induces such decays is
\be
\mathcal{L}_\text{int}\supset \frac{s_hg_D}{4 m_{Z_D}}\left(m_h^2+2m_S^2\right) hSS + s_h g_D m_{Z_D}hZ_D^\mu Z_{D\mu}+ \frac{\lambda_Ss_h}{4}S^3h + \frac{g_D^2s_h}{2}ShZ_{D\mu}Z_{D\mu},
\ee
where the quartic coupling is fixed as $\lambda_S=\frac{1}{2}\left(\frac{g_D m_S}{m_{Z_D}}\right)^2$. From the Lagrangian above, we see that the Higgs can decay to dark photons as $h\to Z_DZ_D$, $h\to SS \to 4Z_D$, $h\to SZ_DZ_D \to 4 Z_D$ and $h \to 3S\to 4Z_D$. Decays with more dark photons in the final state are either suppressed by extra powers of $s_h$, $\epsilon$ or $1/v^2$. We also do not consider intermediate off-shell dark particles. The corresponding 2-body widths are
\bea\label{eq:HiggsInv}
\Gamma(h\to Z_DZ_D)=\left(\frac{g_D s_h}{m_{Z_D}}\right)^2&\frac{1}{32\pi m_h}\left((m_h^2+2m_{Z_D}^2)^2-8(m_h^2-m_{Z_D}^2)m_{Z_D}^2\right)\sqrt{1-\frac{4m_{Z_D}^2}{m_h^2}},\\
\Gamma(h\to SS)&=\left(\frac{g_D s_h}{m_{Z_D}}\right)^2\frac{(m_h^2+2m_S^2)^2}{32\pi m_h}\sqrt{1-\frac{4m_{S}^2}{m_h^2}}.
\eea
Note that in the limit $m_{Z_D},m_S\ll m_h$ both widths are approximately equal
\be\label{eq:hZDZD_approx}
\Gamma(h\to SS)\simeq \Gamma(h\to Z_DZ_D)\simeq \left(\frac{g_D s_h}{m_{Z_D}}\right)^2\frac{m_h^3}{32\pi}.
\ee
The differential 3-body decay widths read
\al{\label{eq:Width_sZDZD}
\frac{\dd \Gamma(h\to SZ_DZ_D)}{\dd m_{12}^2}= \frac{(g_D^2s_h)^2}{256\pi^3}\frac{1}{x_h m_h}\left[2+\left(1-\frac{1}{2x_{Z_D}}\right)^2\right]\sqrt{\lambda\left(1,x_{Z_D},x_{Z_D}\right)\lambda\left(1,x_h,x_S\right)}
}
\al{\label{eq:Width_sss}
\frac{\dd \Gamma(h\to SSS)}{\dd m_{12}^2}= \frac{9(g_D^2s_h)^2}{4096\pi^3}\frac{ 1}{x_h m_h }\frac{x_S^2}{x_{Z_D}^2}\sqrt{\lambda\left(1,x_S,x_S\right)\lambda\left(1,x_h,x_S\right)},
}
with $x_{i}\equiv m_{i}^2/m_{12}^2$ and $\lambda(a,b,c) = a^2+b^2 +c^2 - 2 (ac+ab+bc)$. Furthermore, $m_{12}^2$ takes values in the range $\left[4m_{Z_D}^2,(m_h-m_S)^2\right]$ for eq.~\eqref{eq:Width_sZDZD} and $\left[4m_{S}^2,(m_h-m_S)^2\right]$ for eq.~\eqref{eq:Width_sss}. Taking $m_{Z_D},m_S\ll m_h$ we obtain an approximate expression for the integral of eq.~\eqref{eq:Width_sZDZD}
\be\label{eq:sZDZD_approx}
\Gamma(h\to S Z_DZ_D)\simeq \frac{(g_D^2s_h)^2}{3(64)^2\pi^3}\frac{m_h^5}{ m_{Z_D}^4}\left(1-12 \frac{m_{Z_D}^2}{m_h^2} + 108\frac{m_{Z_D}^4}{m_h^4}\right),
\ee
which, similarly to Eq. \eqref{eq:hZDZD_approx}, is independent of $m_S$. More precisely, the formula above holds for $m_S\lesssim 10$ GeV. For higher dark Higgs masses the relative error between this approximation and the exact result can be larger than $20 \%$.

%
\begin{figure}[t]
\begin{center}
\includegraphics[width=0.48\textwidth]{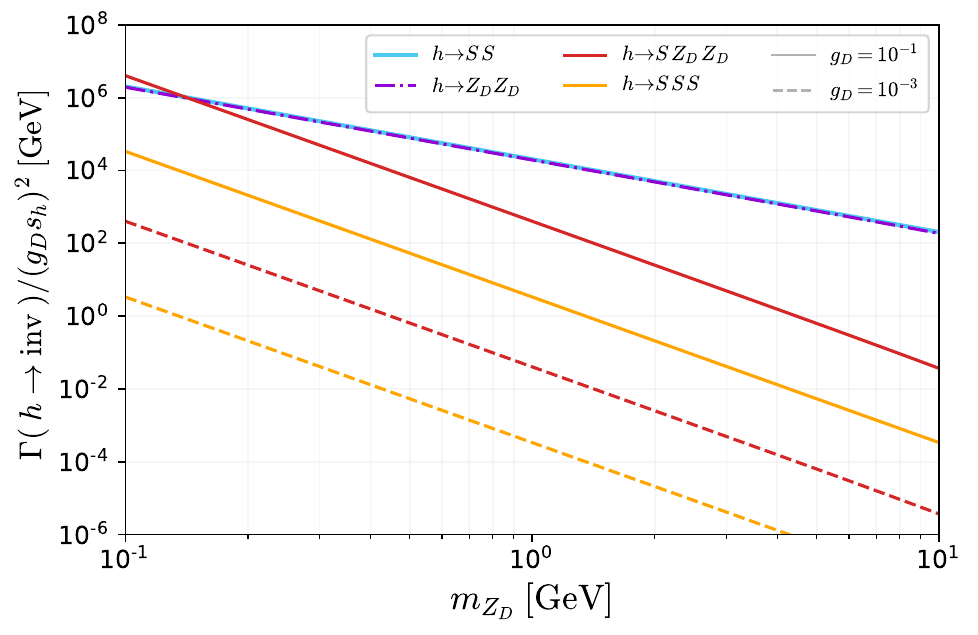}
\includegraphics[width=0.48\textwidth]{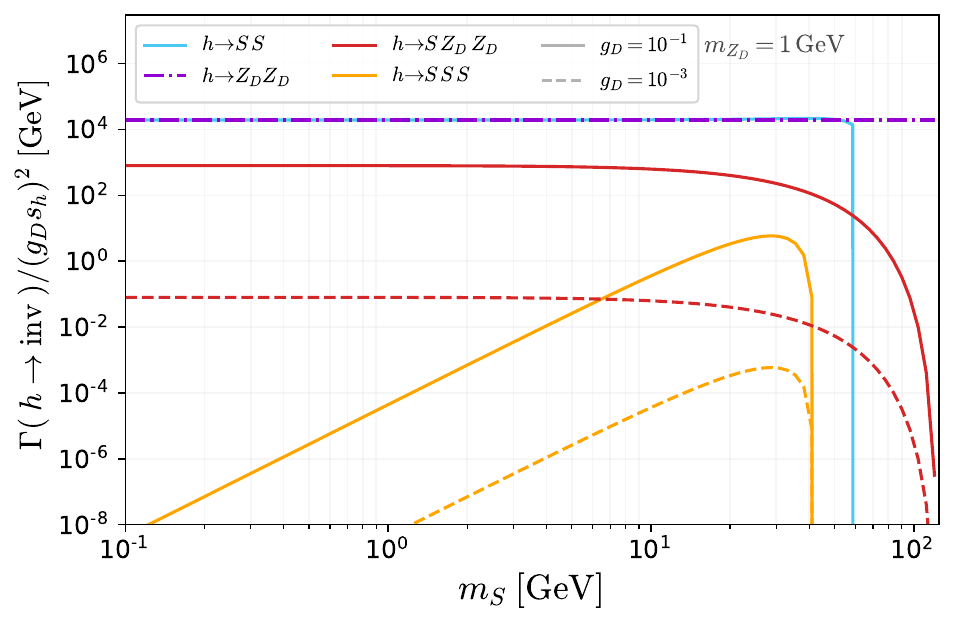}
\end{center}
\vglue -0.8cm
\caption{\label{fig:WidthHinv}
Contributions to the $h$ total decay width 
from $h \to SS$ (blue), $h\to Z_{D} Z_D$ (purple), $h \to S Z_D Z_D$ (red) and $h \to SSS$ (yellow) for $g_D=10^{-1}(10^{-3})$ solid (dashed) lines as a function of $m_{Z_D}$ (left panel) and $m_S$ for $m_{Z_D}= 1$ GeV (right panel).
}
\end{figure}

In figure~\ref{fig:WidthHinv} we compare the contributions to the $h$ total decay width for the four channels described above. The left panel shows the values of the widths, normalized by $(g_D s_h)^{-2}$, as a function of the dark photon mass, considering two values of dark gauge coupling: $g_D = 10^{-1}$ (solid) and $g_D = 10^{-3}$ (dashed). The colors represent the different channels as we indicate in the plot labels. Note that in the case of $h \to SS$ and $h \to Z_D Z_D$ the normalized width does not change for different choices of $s_h$ or $g_D$ due to the $(g_D s_h)^{-2}$ normalization. This is not the case for the 3-body decay channels. The right panel shows the same but as a function of the dark Higgs mass and for a fixed dark photon mass of $m_{Z_D}= 1$ GeV. Note that the 2-body channels are independent of the scalar mass while the $h \to S Z_D Z_D$ is almost independent until $m_S \approx 10$ GeV. We can also see that the $h \to S S S$ mode reach its peak for higher scalar mass values but is always at least two orders of magnitude suppressed in comparison with $h \to S Z_D Z_D$.

\section{Hadronic matrix elements}\label{app:BtoK}

\begin{table}[htb]
\centering
\begin{tabular}{ |p{2cm}|p{4.0cm}|p{3.0cm}|p{2.0cm}|}
\hline
\multicolumn{4}{|c|}{$f_X$}\\
\hline \hline
$X$ & $a_X$  & $b_X$ & $c_X$ [GeV]\\
\hline
$K^\pm$ & 0. & 0.33 & 6.13  \\
$K^{*\pm}(892)$ & 1.36 & -0.99 & 6.06  \\
$K^{*\pm}(1410)$ & $0.22 \left(1- \frac{2 m_{K^*}^{2}}{m_B^2 + m_{K^*}^{2} - m_S^2} \right)$ & $0.28 \; \left(\frac{m_{K^*}}{m_B}\right)$ & $m_B$  \\
$K^{*\pm}(1680)$ & $0.18 \left(1- \frac{2 m_{K^*}^{2}}{m_B^2 + m_{K^*}^{2} - m_S^2} \right)$ & $0.24 \; \left(\frac{m_{K^*}}{m_B}\right)$ & $m_B$  \\
\hline
\end{tabular}
\caption{\label{tab:formfactor1} Values of the coefficients for the form factor $f_X$ used in this work, which were taken from \cite{Ball:2004rg, Ball:2004ye} and  \cite{Lu:2011jm}.
Here, $m_{K^*}$ is the mass of the corresponding vector kaon.}
\end{table}

\begin{table}[htb]
\centering
\begin{tabular}{ |p{2cm}|p{3.0cm}|p{1.0cm}|p{1.0cm}|p{1.0cm}|p{1.0cm}|p{1.0cm}|}
\hline
\multicolumn{7}{|c|}{$A_X$}\\
\hline \hline
$X$ & $g_1(X)$  & $\alpha_1(X)$ & $\beta_1(X)$ & $g_2(X)$  & $\alpha_2(X)$ & $\beta_2(X)$\\
\hline
$K_0(700)$ & 0.46 & 1.6 & 1.35 & 0 & 0 & 0  \\
$K_0(1430)$ & 0.17 & 4.4 & 6.4 & 0 & 0 & 0  \\
$K_1(1270)$ & -0.13 & 2.4 & 1.78 & -0.39 & 1.34 & 0.69  \\
$K_1(1400)$ & 0.17 & 2.4 & 1.78 & -0.24 & 1.34 & 0.69  \\
$K_2(1430)$ & $0.23 \, \left( 1 -\frac{m_S^2}{m_B^2} \right)^{-1}$ & 1.23 & 0.76 & 
0 & 0 &0 \\
\hline
\end{tabular}
\caption{\label{tab:formfactor2} Values of the coefficients for the for factor $A_X$ used 
in this work that were taken from \cite{Sun:2010nv,Bashiry:2009wq} and \cite{Cheng:2010yd}.}
\end{table}

We collect here the expressions for the hadronic elements used in section~\ref{sec:lhcb} that are relevant for $B^\pm \to K S$ decays. For the pseudo-scalar kaons ($K^\pm$) the matrix element can be written as~\cite{Winkler:2018qyg}
\begin{equation}
     \vert \langle K\vert \bar{s}_L b_R\vert B^{\pm}\rangle\vert^2 =     \frac{1}{4} \frac{(m_B^2-m_K^2)^2}{(m_b-m_s)^2} f^2_{K}\, ,
\end{equation}
while for scalar kaons ($K_0^{*\pm} (700), K_0^{*\pm}(1430)$) it takes the form 
\begin{equation}
     \vert \langle K_0\vert \bar{s}_L b_R\vert B^{\pm}\rangle\vert^2 =    \frac{1}{4} \frac{(m_B^2-m_{K_0}^2-m_S)^2}{(m_b+m_s)^2} A^2_{K_0}\,.
\end{equation}
For vector kaons ($K^{*\pm} (892), K^{*\pm}(1410), K^{*\pm}(1680)$) we have instead
\begin{equation}
    \vert \langle K^*\vert \bar{s}_L b_R\vert B^{\pm}\rangle\vert^2 =    \frac{1}{4} \frac{\lambda(m_B^2,m_{K^*}^2,m_S^2)}{(m_b+m_s)^2} f^2_{K^*}\, ,\\
 \end{equation}
and for axial vector kaons ($K_1^\pm (1270), K_1^\pm (1400)$) we obtain
 \begin{equation}
    \vert \langle K_1\vert \bar{s}_L b_R\vert B^{\pm}\rangle\vert^2 =    \frac{1}{4} \frac{\lambda(m_B^2,m_{K_1}^2,m_S^2)}{(m_b-m_s)^2} A^2_{K_1}\,.
 \end{equation}
Finally, in the case of a tensor kaon ($K_2^{*\pm} (1430)$) the matrix element is
  \begin{equation}
    \vert \langle K_2\vert \bar{s}_L b_R\vert B^{\pm}\rangle\vert^2 =    \frac{2}{3} \frac{\lambda^2(m_B^2,m_{K_2}^2,m_S^2)}{(m_b+m_s)^2} \left(\frac{m_B}{m_{K_2}}\right)^2 \, A^2_{K_2}\,.
 \end{equation}
In all expressions above $m_b$ ($m_s$) is the $b$-quark ($s$-quark) mass and  the form factors can be written as~\cite{Winkler:2018qyg,Boiarska:2019jym}
\begin{equation}
f_X =  \displaystyle \frac{a_X}{1-(\frac{m_S}{m_B})^2} +  \frac{b_X}{1-(\frac{m_S}{c_X})^2}\, , 
\end{equation}
\begin{equation}
    A_X  =  \displaystyle \frac{g_1(X)}{1-\alpha_1(X)(\frac{m_S}{m_B})^2 + \beta_1(X) (\frac{m_S}{m_B})^4} +  \frac{g_2(X)}{1-\alpha_2(X)(\frac{m_S}{m_B})^2 + \beta_2(X) (\frac{m_S}{m_B})^4}\, , 
\end{equation}
where the coefficients can be found in Tables~\ref{tab:formfactor1} and ~\ref{tab:formfactor2}.

\bibliographystyle{JHEP2}

\bibliography{HAHM_paper}  

\end{document}